\documentclass[preprint,showpacs,preprintnumbers,amsmath,amssymb]{revtex4}

\usepackage{graphicx}
\usepackage{dcolumn}
\usepackage{bm}

\begin{document}

\preprint{}

\title{Ratios of maximal concurrence-parameterized separability functions, and generalized Peres-Horodecki conditions}
\author{Paul B. Slater}%
\email{slater@kitp.ucsb.edu}
\affiliation{%
ISBER, University of California, Santa Barbara, CA 93106\\
}%
\date{\today}

\begin{abstract}
The probability that a generic real, complex or quaternionic 
two-qubit state is 
separable can be considered to be the sum of three contributions.
One is from those states that are {\it absolutely} separable, that is 
those (which can not be entangled by unitary transformations) for which the {\it maximal concurrence} over spectral orbits 
($C_{max}$) is zero. The other two contributions are from the states
for which $C_{max} \in (0,\frac{1}{2}]$, and for which
 $C_{max} \in [\frac{1}{2},1]$. We have previously (arXiv:0805.0267) found exact formulas for the absolutely separable contributions in terms of the Hilbert-Schmidt metric over the quantum states, and here advance hypotheses as to the exact contributions for  $C_{max} \in [\frac{1}{2},1]$. A crucial element 
in understanding the two contributions for $C_{max}>0$ is the nature of the ratio 
($R$) of the $C_{max}$-parameterized separability function for the complex states to the {\it square} of the comparable function for the real 
states--both such functions having clearly displayed jump discontinuities at 
$C_{max}=\frac{1}{2}$. For  $C_{max} \in (0,\frac{1}{2}]$, the ratio $R$ appears to be of the form $1+ k C_{max}$, except near $C_{max}=\frac{1}{2}$, while for $C_{max} \in [\frac{1}{2},1]$, there is strong numerical evidence that it equals 2 (thus, according to the Dyson-index pattern of random matrix theory). Related phenomena also occur for the 
minimally-degenerate two-qubit states and the qubit-qutrit states. Our results have immediate application to the computation of separability probabilities in terms of other metrics, such as the Bures (minimal monotone) metric. The paper begins with continuous embeddings of the
separability probability question in terms of four metrics of 
interest, using "generalized Peres-Horodecki conditions".

\end{abstract}


\maketitle

\tableofcontents
\section{Introduction}
One possibly productive strategy to pursue when confronted with 
an apparently intractable problem, is to embed it in some broader
class of problems. Doing so, 
hopefully, may lead to new insights and progress, 
including ones regarding the original (smaller) problem. 
In the first of the two basic parts of this paper (secs.~\ref{secqubqub}-\ref{Remarks}), 
we adopt  such embedding strategies 
for the task of determining the probabilities--with regard to a 
number 
of metrics of quantum-mechanical interest--that certain generic forms of 
$2 \times 2$ or $2 \times 3$  quantum system are separable 
\cite{ZHSL,slaterA,slaterC,slaterJGP,slaterPRA,pbsCanosa,slater833,JMP2008}.
(As computers presumably grow more powerful, these readily-formulated, 
but high-dimensional [9, 15, \ldots] and high-degree [{\it e. g.}, quartic] problems may eventually lose their apparent 
present-day intractability (cf. \cite{Ioannou,Ye})--much as did the
famous four-color planar map theorem of Appel and Haken \cite{appel,soifer}. Nevertheless, 
it would certainly be 
appealing to address these problems with more theoretical understanding than is required by "brute force" 
computation (cf. \cite{szHS,szBures,andai,ingemarkarol}).)

In the second basic part (secs.~\ref{secpiecewise}-\ref{Remarks2}), 
building upon our recent work in \cite{maxconcur4}, we attempt to gain insight--using manifest relations to random matrix theory--into the very same separability probability questions by determining the nature of certain eigenvalue-parameterized separability functions. These are expressed as univariate functions of the maximal concurrence over spectral orbits.

The fundamental question being addressed here of determining the probability that a generic bipartite
quantum state is separable or not was first raised by {\.Z}yczkowski, Horodecki, Sanpera and Lewenstein in a pioneering, much-cited 1998 paper 
\cite{ZHSL}. As motivation they wrote: "One of the fundamental questions concerning these subjects is to estimate how many entangled (disentangled) states 
exist among all quantum states. More precisely, one can consider the problem of quantum separability or inseparability 
from a measurement theoretical point of view, and ask about 
relative volumes of both sets. There are three main reasons 
of importance in this problem. The first reason, of some 
philosophical implication, may be contained in the questions 
ÔÔIs the world more classical or more quantum? Does it contain more quantum-correlated (entangled) states than classically correlated ones?ÕÕ The second reason has a more practical origin. Analyzing some features of entanglement, one often has to rely on numerical simulations. It is then important to know to what extent entangled quantum states may be 
considered as typical. Finally, the third reason has a physical 
origin. The physical meaning of separability has recently 
been associated with the possibility of partial time reversal" \cite[p. 883]{ZHSL}.

In sec.~\ref{secqubqub}, in the first basic part of the paper, we analyze the cases 
of  generic 9-dimensional real 
and 15-dimensional complex two-qubit systems. 
For our calculations, we utilize the Euler-angle parameterizations
of the real (developed by S. Cacciatori 
\cite[App. A]{JMP2008}) and of the complex $4 \times 4$ 
density matrices ($\rho$) \cite{tbs}, as well as 
the Tezuka-Faure (TF) procedure \cite{tezuka,giray1} 
for generating low-discrepancy sets of (9- and 15-dimensional) points. 
These points are employed for quasi-Monte Carlo numerical integration 
with respect to the {\it product} of the (6- or 12-dimensional) Haar measure over the Euler angles and 
(3-dimensional) metric-specific 
measures over the eigenvalues of the density matrices.
In sec.~\ref{Hyperarea},
we turn our attention to parallel "continuous embedding" analyses 
pertaining to 
the 14-dimensional (rank-3) boundary of 
the 15-dimensional generic complex $4 \times 4$ density matrices.

In sec.~\ref{secqubqut}, we 
make use of the $SU(6)$-based Euler-angle parameterization
of the 35-dimensional generic complex qubit-{\it qutrit} $6 \times 6$ density matrices 
\cite[sec. XI]{sudarshan} to investigate the corresponding
rank-6 and rank-5 
problems. In sec.~\ref{extended}, we investigate formally
extending the range of our basic  parameter ($\alpha$)--used in forming
convex combinations--from beyond 
[0,1] to $[-\infty,\infty]$.
In sec.~\ref{secconcurrence}, we 
depart from our initial paradigm, employing the 
parameter $\alpha$, and evaluate 
the separability probabilities of the generic complex and 
real two-qubit states for which 
the entanglement measure 
{\it concurrence} ($C$) \cite{wkw,iwai} is less than
some threshold. 
(There we observe some interesting behavior, involving the 
{\it intersection} of the curves  for different metrics. 
For $C=1$, we obtain the usual separability probabilities.) 

The concept of {\it maximal} concurrence ($C_{max}$) over spectral orbits \cite{roland2}
\begin{equation} \label{maximalconcurrenceformula}
C_{max}=\mbox{max}(0,\lambda_1-\lambda_3 
-2 \sqrt{\lambda_2 \lambda_4}), \hspace{.5in}
(\lambda_1 \geq \lambda_2 \geq \lambda_3 \geq \lambda_4)
\end{equation}
(a quantity which can not be increased under unitary transformations) of a two-qubit density matrix ($\rho$), where the $\lambda$'s are the ordered eigenvalues of $\rho$,
is used in the second 
basic set of analyses of the paper (sec.~\ref{secpiecewise}). There, we importantly add to certain findings \cite{maxconcur4} concerning the strong goodness-of-fit to two-qubit
eigenvalue-parameterized separability functions (ESFs) of piecewise functions
of $C_{max}$. We find evidence
of adherence over a {\it half}-domain $C_{max} \in [\frac{1}{2},1]$ to a Dyson-index pattern {\it both} for the 
generic rank-4 (first investigated in \cite{maxconcur4}) and generic
rank-3 (as found here [sec.~\ref{rankthreesub}]) real and complex two-qubit states. We, further, undertake an analogous examination of: (a) 
the generic rank-5 qubit-qutrit states in sec.~\ref{rankfive}, observing 
interesting jump discontinuities in the ESFs; and (b) the generic full rank qubit-qutrit states in sec.~\ref{ranksix}, where, again, the Dyson-index
pattern appears to emerge over a restricted domain  $C_{max} \in 
[\frac{1}{3},1]$.  Additionally, in  
sec.~\ref{CmaxDomain}, we are able to present
new simple exact results pertaining to certain components of the 
desired Hilbert-Schmidt separability probabilities. In these 
regards, let us draw the 
reader's attention, particularly, to (the titular) Figs.~\ref{fig:Convenience2} 
and \ref{fig:ConvenienceArea}

Since when we had earlier addressed  the issue of two-qubit separability probabilities in terms of {\it diagonal-entry}-parameterized separability functions (DESFs) \cite{slater833}, we found apparently {\it total} agreement with Dyson-index behavior, the need (remaining unmet)
to reconcile these two forms of Dyson-index patterns (full and partial) is obvious.

Remarks relevant to the two primary 
sets of analyses--which share the use of 
concurrence and are devoted to the determination of  separability {\it probabilities}--are given in secs.~\ref{Remarks} and \ref{Remarks2}.
\subsection{Separability functions}
Let us state here that the concept of a {\it separability function}--both in its eigenvalue-parameterized (ESF) and diagonal-entry-parameterized (DESF) forms--has been developed in order to reduce the intrinsically high dimensionalities of the generic separability probability questions. By integrating over the majority of parameters--for example, Euler angles or off-diagonal entries--one reduces--the problems 
(at least, in the two-qubit case) to ones of (only!) a three-dimensional nature. It also appears possible to further reduce the three-dimensional problems to single-dimensional ones by finding an appropriate parameter--known to be the ratio of the product 
of the 11- and 44-diagonal entries to the product of the 22- and 33-diagonal entries in the DESF-case, and apparently (as numerics strongly indicate) the maximal concurrence over spectral orbits in the ESF-case.
\subsection{Metrics employed}
The metrics of quantum-mechanical interest that we utilize to form 
({\it via} their Riemannian volume elements) measures over the quantum
systems are
the (Euclidean or flat, non-monotone \cite{ozawa}) Hilbert-Schmidt 
\cite{szHS}, and three monotone metrics 
\cite{petzsudar}--the (minimal monotone) 
Bures \cite{szBures}, Wigner-Yanase \cite{wigneryanase} and 
Kubo-Mori \cite{petz1994} metrics. (We also attempted to include 
the information-theoretically significant monotone "quasi-Bures" 
[Grosse-Krattenthaler-Slater] 
metric \cite{krattenthaler,slaterPRA,hayashisan}, 
which yields the minimax/maximin asymptotic 
redundancy for universal quantum coding,
but encountered
some initial, at least, numerical difficulties in this regard.)
\subsection{Prior conjectures}
In \cite{slater833}, we were led by a combination of numerical and
theoretical [Dyson-index-related] arguments, involving 
{\it diagonal-entry}-parameterized separability functions (DESFs),  to conjecture that the Hilbert-Schmidt separability probabilities are, respectively, 
$\frac{8}{33} \approx  0.242424$ for the generic complex two-qubit systems and  $\frac{8}{17} \approx 0.470588 $ for its real counterpart. 
(The supporting evidence appeared particularly strong for 
the $\frac{8}{33}$ figure.) Also, it has been  
further conjectured that for the generic {\it quaternionic} two-qubit
systems, the corresponding probability is 
$\frac{72442944}{936239725} \approx 0.0733389$
\cite[eq. (15)]{slaterJGP2} \cite[p. 25]{JMP2008}).

We had also earlier advanced in \cite[Table VI]{slaterJGP},  the
"silver mean" (that is, $\sqrt{2}-1$) 
conjectures that the generic complex two-qubit 
{\it Bures} separability probability is 
\begin{equation} \label{silvermean}
P_{sep/Bures}^{complex} =\frac{1680 (\sqrt{2}-1)}{\pi ^8} \approx 0.0733389.
\end{equation}
and the corresponding Kubo-Mori analogue is
\begin{equation}
P_{sep/KM}^{complex}= \frac{1575 (\sqrt{2} -1)}{2 \pi ^8}
\approx 0.035398.
\end{equation}
(Additionally, for the "average monotone metric"--not employed in 
this paper-it 
was conjectured in \cite{slaterJGP} that the associated separability probability is 
$\frac{81664 (\sqrt{2} -1)}{75 \pi ^8} 
\approx 0.0475329$. Further still, the Wigner-Yanase separability probability
was hypothesized to equal the ratio of $\frac{7 (\sqrt{2}-1)}{4}$ to the not-yet-determined
Wigner-Yanase volume of the generic [entangled and separable] complex two-qubit states.)

In \cite[sec. X]{slater833}, again studying the corresponding DESFs, the conjectures were put forth that the generic real and complex
qubit-{\it qutrit} Hilbert-Schmidt separability probabilities are, $\frac{32}{213} \approx 0.150235$ and 
(agreeing very closely with the numerics) $\frac{32}{1199} \approx 
0.0266889$, 
respectively.

\section{Generic full-rank real and complex two-qubit cases} \label{secqubqub}
\subsection{First set of constraints--convex combinations of determinants of $\rho$ and $\rho_{PT}$}
\subsubsection{Real two-qubit density matrices}
\begin{figure}
\includegraphics{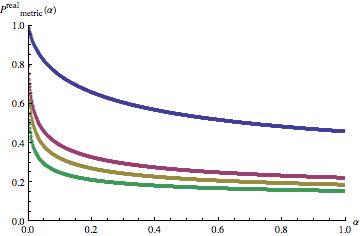}
\caption{\label{fig:real1}Curves generated by enforcement of 
determinant-based constraint (\ref{constraint1}) for the generic 9-dimensional real density matrices. The order of dominance of the four metric-specific curves is
given in (\ref{order}). 17,100,000 Tezuka-Faure 9-dimensional points
were employed in the quasi-Monte Carlo numerical integration. 
The values at $\alpha=1$ are the embedded (conventional) separability probabilities.}
\end{figure}

In Fig.~\ref{fig:real1} we show as a function of $\alpha \in [0,1]$ the
probabilities ($P^{real}_{metric}(\alpha)$), 
in terms of the four metrics under
consideration, that for a generic (9-dimensional) real two-qubit system 
\begin{equation} \label{constraint1}
\alpha |\rho_{PT}| +(1-\alpha) |\rho| \geq 0.
\end{equation}
Here $\rho_{PT}$ is the partial transpose of $\rho$ and $|\rho_{PT}|$, its 
determinant. Of course, here and throughout we incorporate into our analyses, 
the original, notable Peres-Horodecki 
necessary and sufficient conditions for separability in terms of the nonnegativity
of $\rho_{PT}$ \cite{asher,michal}. The partial transpose of a $4 \times 4$ density matrix can have at most one negative eigenvalue, so the condition
$|\rho_{PT}| < 0 $ is fully equivalent to  $\rho_{PT}$ having a single negative eigenvalue \cite{augusiak}. (Also, obviously, the nonnegativity 
condition $|\rho|  \geq 0$ is always satisfied.)

The order of dominance of the 
four monotonically-decreasing curves 
in Fig.~\ref{fig:real1}, as well as all the other analogous curves below, 
turns out--with the important exception of those in sec.~\ref{both}, where 
we observe {\it intersecting} behavior--to be
\begin{equation} \label{order}
\mbox{Hilbert-Schmidt} > \mbox{Bures} > \mbox{Wigner-Yanase} > \mbox{Kubo-Mori}.
\end{equation}
This, of course, will imply that the associated two-qubit 
{\it separability} probabilities (corresponding to $\alpha=1$)
adhere to the same ordering.
Since the Bures metric is also the {\it minimal} monotone metric, it is not surprising
that it is extremal among the three monotone metrics under consideration. 
In estimating these curves, as well as all others displayed below 
involving $\alpha$--except Fig.~\ref{fig:elongated}--we subdivided 
the unit interval $\alpha \in [0,1]$ into one thousand subintervals.

We can fit the Hilbert-Schmidt curve in Fig.~\ref{fig:real1} rather 
well--the integral over $\alpha \in [0,1]$ of the sum of squares of the differences being only 0.00026557--while exactly 
achieving the conjectured separability probability of $\frac{8}{17}$,
with the simple function
\begin{equation}\label{fitReal}
P^{real}_{HS}(\alpha)= \frac{8}{8+9 \sqrt{\alpha}}.
\end{equation}

\subsubsection{Complex two-qubit density matrices}
\begin{figure}
\includegraphics{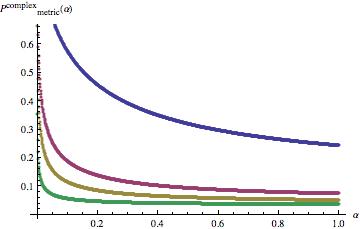}
\caption{\label{fig:complex1}Curves generated by enforcement of 
determinant-based constraint (\ref{constraint1}) for the generic 15-dimensional complex density matrices. 
The order of dominance of the four metric-specific curves is
given in (\ref{order}). 15,400,000 Tezuka-Faure 15-dimensional points
were employed in the numerical integration.}
\end{figure}

In Fig.~\ref{fig:complex1} we 
analogously show as a function of $\alpha \in [0,1]$, the four 
probabilities ($P^{complex}_{metric}(\alpha)$) for a generic (15-dimensional) complex two-qubit system that 
the inequality constraint (\ref{constraint1}) is satisfied. 
We can fit the Hilbert-Schmidt 
curve here very well--the integral over $\alpha \in [0,1]$ of the sum of squares of the differences being only 0.00030092--while achieving our conjectured
separability probability of $\frac{8}{33}$ \cite{slater833} with the function
\begin{equation}
P^{complex}_{HS}(\alpha) = (\frac{c}{c+ 25 \sqrt{\alpha}})^2,
\end{equation}
where $c=8+2 \sqrt{66}$. 
Also, the {\it square} of 
the real counterpart (\ref{fitReal}) does provide a close fit to the Hilbert-Schmidt curve in Fig.~\ref{fig:complex1}. (However, our conjectured
complex two-qubit separability probability of $\frac{8}{33} \approx 0.242424$ is not equal to the square, $\frac{64}{289} \approx 0.221453$, of 
the conjectured real separability probability of $\frac{8}{17}$, so conformity to a Dyson-index pattern is not total.)

Additionally, we can very well fit the Bures curve in Fig.~\ref{fig:complex1} {\it and} our corresponding conjectured 
"silver mean" separability 
probability (\ref{silvermean}) \cite{slaterJGP} by the function (of the seventh root of $\alpha$)
\begin{equation} \label{conjBures}
P^{complex}_{Bures}(\alpha) =\frac{1680 (1-\sqrt{2})}{\left(4 \sqrt{105
   \left(-1+\sqrt{2}\right)} \left(\sqrt[7]{\alpha }-1\right)+\pi ^4
   \sqrt[7]{\alpha }\right)^2}.
\end{equation}
(The sum-of-squares measure of fit between the two curves 
is 0.000739208. However, the  [exact] square root of 
(\ref{conjBures})--deviating substantially from a Dyson-index-like pattern--does not at all provide a close fit [as in the HS case] 
to the real {\it Bures} counterpart in Fig.~\ref{fig:real1}.)
\subsection{Second set of constraints--convex combinations of minimum eigenvalues of $\rho$ and $\rho_{PT}$}
\subsubsection{Real two-qubit density matrices}
\begin{figure}
\includegraphics{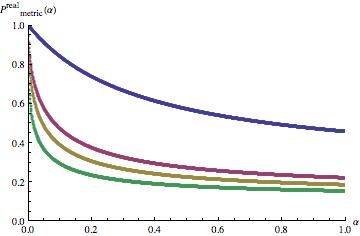}
\caption{\label{fig:real2}Curves generated by enforcement of 
minimum-eigenvalue-based constraint (\ref{constraint2}) for the generic 9-dimensional real density matrices. The order of dominance of curves is
given in (\ref{order}). 32,700,000 Tezuka-Faure 9-dimensional points
were employed in the numerical integration.}
\end{figure}

In Fig.~\ref{fig:real2} we show as a function of $\alpha \in [0,1]$, the
probabilities for a generic (9-dimensional) real two-qubit system that
\begin{equation} \label{constraint2}
\alpha {\lambda_{PT}}_{min} +(1-\alpha) \lambda_{min} \geq 0,
\end{equation}
where the subscript $min$ denotes the smallest of the corresponding
four eigenvalues ($\lambda$) of either
$\rho$ or $\rho_{PT}$. (As noted, having all  eigenvalues nonnegative is fully
equivalent to having a nonnegative determinant for the partial transpose
of a $4 \times 4$ density matrix \cite{augusiak}. The entanglement measure {\it negativity} 
is equal to $\mbox{max}[0,-2 \lambda_{PT_{min}}]$ \cite[p. 401]{ingemarkarol}. The Hilbert-Schmidt distance of an entangled state to the set of all partially transposed sets can be expressed as a function of 
the negative eigenvalues of the partial transpose of the entangled state 
\cite{verdehmoor}.)
\subsubsection{Complex two-qubit density matrices}
\begin{figure}
\includegraphics{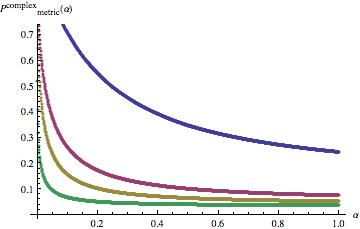}
\caption{\label{fig:complex2}Curves generated by enforcement of 
minimum-eigenvalue-based 
constraint (\ref{constraint2}) for the generic 15-dimensional complex density matrices. The order of dominance of curves is
given in (\ref{order}). 18,450,000 Tezuka-Faure 15-dimensional points
were employed.}
\end{figure}

In Fig.~\ref{fig:complex2} we show as a function of $\alpha \in [0,1]$, the
probabilities for a generic (15-dimensional) complex two-qubit system that 
the inequality constraint (\ref{constraint2}) holds.
One can fit the complex HS curve rather closely by the {\it square} of
the corresponding real HS curve, particularly so if one adds a small linearly increasing 
correction of the form $\frac{\it \alpha}{30}$ to this square. 
\subsection{Third set of constraints--determinants of convex combinations
of $\rho$ and $\rho_{PT}$}
\subsubsection{Real two-qubit density matrices}
In Fig.~\ref{fig:real3} we show as a function of $\alpha \in [0,1]$, the
probabilities for a generic (9-dimensional) real two-qubit system that 
the positive "twofold partial" transpose condition,
\begin{equation} \label{constraint3}
| \alpha \rho_{PT} +(1-\alpha) \rho| \geq 0,
\end{equation}
holds.
The Hilbert-Schmidt curve is highly linear in character. 
The line $1-\frac{9 \alpha }{17}$ closely approximates it, as well
as reproducing the conjectured separability probability of $\frac{8}{17}$.
\subsubsection{Complex two-qubit density matrices}
In Fig.~\ref{fig:complex3} we show as a function of $\alpha \in [0,1]$, the four 
probabilities ($P^{complex}_{metric}(\alpha)$) for a generic 
complex two-qubit system that 
the inequality (\ref{constraint2}) holds.

\begin{figure}
\includegraphics{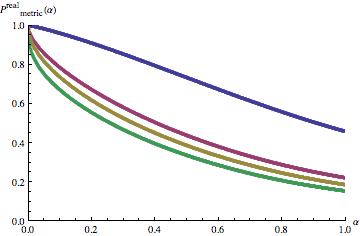}
\caption{\label{fig:real3}Curves generated by enforcement of determinant constraint (\ref{constraint3}) for the generic 9-dimensional real density matrices. The order of dominance of curves is
given in (\ref{order}). 22,500,000 TF 9-dimensional points
were employed.}
\end{figure}

\begin{figure}
\includegraphics{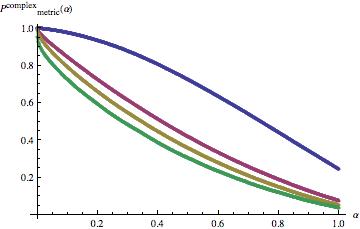}
\caption{\label{fig:complex3}Curves generated by enforcement of determinant constraint (\ref{constraint3}) for the generic 15-dimensional complex density matrices. The order of dominance of curves is
given in (\ref{order}). 14,520,000 TF 15-dimensional points
were employed.}
\end{figure}

In computing Figs.~\ref{fig:real3} and \ref{fig:complex3}, we solve 
the quartic equation $| \alpha \rho_{PT} +(1-\alpha) \rho| =0$ and
assume
that there can not be more
than one solution $\alpha \in [0,1]$. (Numerically, this did appear to be the case, except
for some isolated instances in which two positive [essentially identical] roots both very close to zero were found.)

Another possible generalized Peres-Horodecki condition--that is, that the minimum eigenvalue
of $\alpha \rho_{PT} +(1-\alpha) \rho$ be nonnegative--appeared to be considerably more
problematical (time-consuming) 
than (\ref{constraint3}) to investigate, though we do, in fact, implement such a condition for the generic complex qubit-qutrit systems (sec.~\ref{secqubqut}) and in generating Fig.~\ref{fig:elongated}.

Let us note that all the curves displayed 
so far in this communication appear to
correspond to {\it convex} 
functions, but for the last two Hilbert-Schmidt curves.
\subsection{Comparison of metric-specific curves for the first three sets of constraints}
\begin{figure}
\includegraphics{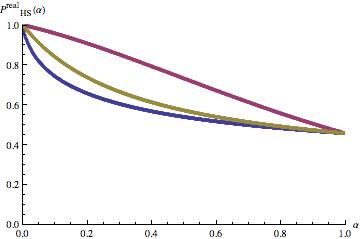}
\caption{\label{fig:crossRealHS}The curves corresponding to the Hilbert-Schmidt metric 
plotted above for the three {\it different} constraints for the generic real 9-dimensional
two-qubit systems. The quasi-linear 
curve based on constraint (\ref{constraint3}) dominates that based on (\ref{constraint2}), which dominates that based 
on (\ref{constraint1}).}
\end{figure}

\begin{figure}
\includegraphics{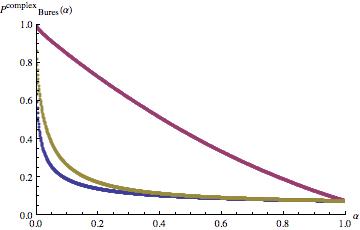}
\caption{\label{fig:crossComplexBures}The curves corresponding to the Bures metric 
plotted above for the three {\it different} constraints for the generic complex 15-dimensional
two-qubit systems. The curve based on constraint (\ref{constraint3}) dominates that based on (\ref{constraint2}), which dominates that based 
on (\ref{constraint1}).}
\end{figure}

In Fig.~\ref{fig:crossRealHS} we show in a single plot, the three
Hilbert-Schmidt curves plotted above (one per figure) for the generic real 9-dimensional
two-qubit systems, while 
in Fig.~\ref{fig:crossComplexBures} we show in a single plot, the three
Bures  curves plotted above for the generic complex 15-dimensional
two-qubit systems. In both these 
figures the order of dominance of the curves
is the same--the curve based on 
the constraint (\ref{constraint3}) dominates that based on
(\ref{constraint2}), which, in turn, dominates that based on (\ref{constraint1}).

Of course, we find in both of these figures that the three curves 
have common points-of-intersection
at  $\alpha=1$, corresponding to the "ordinary" 
separability probability (as well as $\alpha=0$).
\section{Generic rank-3 complex two-qubit case} \label{Hyperarea}
\begin{figure}
\includegraphics{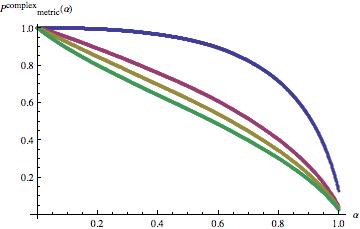}
\caption{\label{fig:Hyperarea1}Curves generated by enforcement of 
{\it modified} ($3 \times 3$ principal minor)
constraint (\ref{constraint1}) for the generic 14-dimensional rank-3 complex density matrices. The order of dominance of curves is
given in (\ref{order}). 7,300,000 TF 14-dimensional points
were employed.}
\end{figure}

\begin{figure}
\includegraphics{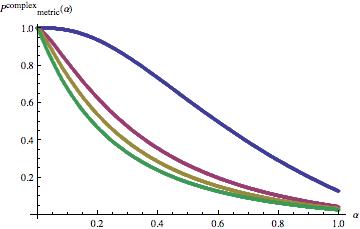}
\caption{\label{fig:Hyperarea3}Curves generated by enforcement of 
{\it modified} (minimum-eigenvalue-based) constraint (\ref{constraint2}) for the generic 14-dimensional rank-3 complex density matrices. The order of dominance of curves is
given in (\ref{order}). 10,300,000 TF 14-dimensional points
were employed.}
\end{figure}

\begin{figure}
\includegraphics{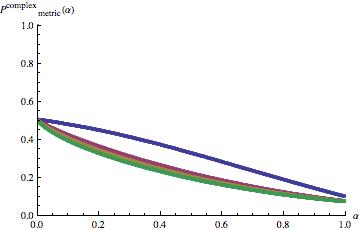}
\caption{\label{fig:Hyperarea2}Curves generated by enforcement of constraint (\ref{constraint3}) for the generic 14-dimensional rank-3 complex density matrices. The order of dominance of curves is
given in (\ref{order}). 4,300,000 TF 14-dimensional points
were employed.}
\end{figure}

The "twofold" volume-to-area-ratio 
theorem of Szarek, Bengtsson and {\.Z}yczkowski 
\cite{sbz} allows us to immediately extend our conjecture \cite{slater833} of Hilbert-Schmidt
separability probability of $\frac{8}{33}$ for the 15-dimensional generic complex 
two-qubit states to the fully equivalent conjecture that the HS separability 
probability of the states on the 14-dimensional boundary is {\it one-half} of this, that is, $\frac{4}{33}$.
In Figs.~\ref{fig:Hyperarea1} and \ref{fig:Hyperarea3} we show our corresponding estimation
of the $\alpha$-separabilities based on certain 
obvious modifications of the determinant constraint 
(\ref{constraint1}) and the minimum eigenvalue
constraint (\ref{constraint2}). 
That is, rather than using the (zero) determinant of the rank-three
density matrix, we use its generically nonzero $3 \times 3$ 
principal minor. Further, rather than using
the minimum (zero) eigenvalue, we employ the minimum of the generically three nonnegative eigenvalues.
 In Fig.~\ref{fig:Hyperarea2} we display the rank-3 
$\alpha$-separability probability estimates based on the application
of the constraint (\ref{constraint3}). Here we notice some unusual
behavior near $\alpha=0$ due to the degeneracy (zero determinant)
of a rank-3 two-qubit ($4 \times 4$) density matrix.

\section{Generic full-rank complex qubit-qutrit case} \label{secqubqut}
\begin{figure}
\includegraphics{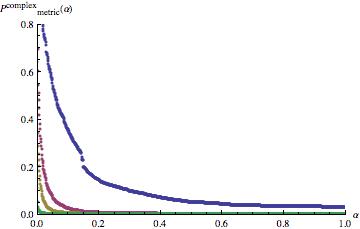}
\caption{\label{fig:complex4}Curves generated by enforcement of 
(minimum-eigenvalue-based) 
constraint (\ref{constraint2}) for the generic 35-dimensional 
complex $6 \times 6$ density matrices. The order of dominance of curves is
given in (\ref{order}). 30,650,000 TF 35-dimensional points
were employed.}
\end{figure}

Of the three distinct sets of constraints considered in the two-qubit case,
only (\ref{constraint2}) seemed immediately adoptable to the qubit-qutrit
case associated with $6 \times 6$ density matrices.
In our computations, we now employ the associated $SU(6)$ Euler-angle parameterization 
\cite[sec XI]{sudarshan}. In Fig.~\ref{fig:complex4} we show the corresponding plot.

\begin{figure}
\includegraphics{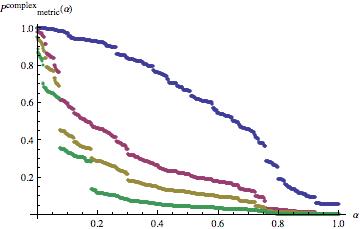}
\caption{\label{fig:complex4b}Curves generated by verifying the nonnegativity of the minimum eigenvalue of 
$\alpha \rho_{PT} +(1-\alpha) \rho$ for the generic complex 
qubit-qutrit systems. The order of dominance of curves is
given in (\ref{order}). 1,300,000 TF 35-dimensional points
were employed.}
\end{figure}

Further, by specifically checking nonnegativity at each value of 
$\alpha =\frac{1}{1000},\dots 1$, we were able to enforce the condition that the minimum eigenvalue of the matrix convex combination
 $\alpha \rho_{PT} +(1-\alpha) \rho$ be nonnegative. (Nonnegativity of the determinant of the partial transpose is no longer equivalent--as it is in the two-qubit case 
\cite{augusiak}--to having no negative eigenvalues, since {\it two} negative 
eigenvalues yields a {\it positive} determinant.) 
The corresponding
plot is displayed in Fig.~\ref{fig:complex4b}.

We can "sandwich" the Hilbert-Schmidt curve in Fig.~\ref{fig:complex4}
between two curves, corresponding to exact squares, both of which yield the conjectured
separability probability of $\frac{32}{1199}$. These functions are
\begin{equation}
\frac{64}{\left(\left(-8+\sqrt{2398}\right) \sqrt{\alpha }+8\right)^2} \hspace{.5in} 
\mbox{and} \hspace{.5in} \frac{64}{\left(\left(-8+\sqrt{2398}\right) \alpha +8\right)^2}.
\end{equation}
\section{Extending range of $\alpha$-parameter} \label{extended}
\begin{figure}
\includegraphics{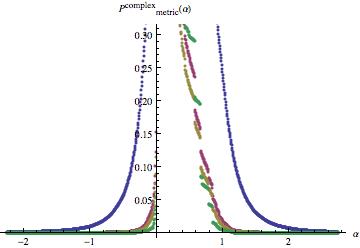}
\caption{\label{fig:elongated}Curves generated by verifying the nonnegativity of the minimum eigenvalue of 
$\alpha \rho_{PT} +(1-\alpha) \rho$ for the generic complex 
two-qubit systems, using 
the {\it extended} interval $\alpha \in [-\frac{9}{4},\frac{11}{4}]$. The order of dominance of curves is
given in (\ref{order}). 1,650,000 TF 15-dimensional points
were employed.}
\end{figure}

We have, so far, considered our primary variable $\alpha$ as extending
over the unit interval [0,1]. However, it appears quite interesting and 
possibly more natural to formally view its range as the real line $[-\infty,\infty]$.
In a further 
analysis, we developed a plot (Fig.~\ref{fig:elongated}) over $\alpha \in[-\frac{9}{4},\frac{11}{4}]$, 
of the estimated 
$\alpha$-probability that 
the $4 \times 4$ matrix $\alpha \rho_{PT} +(1-\alpha) \rho$, where 
$\rho$ is a generic complex two-qubit density matrix, has all 
its four eigenvalues nonnegative.
\section{Concurrence-related analyses} \label{secconcurrence}
\subsection{Generalized Peres-Horodecki conditions}
\begin{figure}
\includegraphics{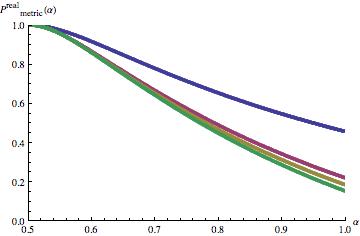}
\caption{\label{fig:realConcur}Implementation of 
concurrence-based constraint (\ref{constraintConcur}) for the generic
9-dimensional real two-qubit states. The order of dominance of curves is
given in (\ref{order}). 9,250,000 TF-points were employed.}
\end{figure}

\begin{figure}
\includegraphics{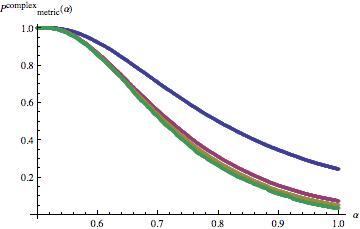}
\caption{\label{fig:complexConcur}Implementation of 
concurrence-based constraint (\ref{constraintConcur}) for the generic
15-dimensional complex two-qubit states. The order of dominance of curves is
given in (\ref{order}). 6,100,000 TF-points were employed.}
\end{figure}

\begin{figure}
\includegraphics{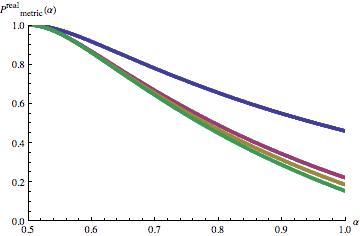}
\caption{\label{fig:realConcurArea}Implementation of 
concurrence-based constraint 
(\ref{constraintConcur}) for the generic
8-dimensional real rank-3 two-qubit states. 10,300,000 TF-points were employed.}
\end{figure}

\begin{figure}
\includegraphics{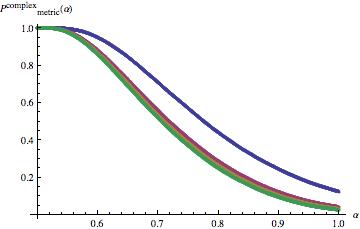}
\caption{\label{fig:complexConcurArea}Implementation of 
concurrence-based constraint 
(\ref{constraintConcur}) for the generic
14-dimensional complex rank-3 two-qubit states. 11,950,000 TF-points were employed.}
\end{figure}

In all the analyses reported above,
the nonnegativity convex combination constraints 
("generalized Peres-Horodecki conditions") utilized, have been
expressed either in terms of the determinant
or the minimum eigenvalue of $\rho_{PT}$. 
In the two-qubit case, we have also been able to investigate
similiarly-motivated conditions using, in conjunction, the {\it maximal concurrence} 
over spectral orbits (\ref{maximalconcurrenceformula}) \cite{roland2} 
of a two-qubit density matrix ($\rho$), and its 
{\it concurrence} \cite{wkw}
\begin{equation} \label{maxcon}
C=\mbox{max}(0,\eta_1-\eta_2 -\eta_3 -\eta_4), \hspace{.5in}  (\eta_1 \geq \eta_2 \geq \eta_3 \geq \eta_4).
\end{equation}
(Here, the $\lambda$'s are the ordered eigenvalues of $\rho$ and the $\eta$'s are the ordered
eigenvalues of $\sqrt{\sqrt{\rho} \tilde{\rho} \sqrt{\rho}}$, 
where $\tilde{\rho} =(\sigma_y \otimes \sigma_y) \rho^{*} 
(\sigma_y \otimes \sigma_y)$, and $\sigma_y \equiv \sigma_2$ 
is a Pauli matrix, and $*$ denotes conjugation. 
Throughout the reminder of the paper, the symbol $\sigma$--consistently with our previous notation--will denote a "separability function, and not a Pauli matrix.)
The corresponding constraint we employ is
\begin{equation} \label{constraintConcur}
- \alpha C +(1-\alpha) C_{max} \geq 0.
\end{equation}
Since $C_{max} \geq C$, the constraint 
holds trivially for $\alpha \in [0,\frac{1}{2}]$.
In Figs.~\ref{fig:realConcur} and \ref{fig:complexConcur}, 
we show for the {\it half}-interval 
$\alpha \in [\frac{1}{2},1]$, the curves for the generic real and complex two-qubit states, respectively, based on (\ref{constraintConcur}), while in 
Figs.~\ref{fig:realConcurArea} and \ref{fig:complexConcurArea}, we
display the corresponding plots for the 
generic rank-3 real and complex two-qubit states, respectively.
\subsection{Separability probabilities as functions of concurrence--intersecting curves} \label{both}
In this  section, we depart from the basic paradigm so far employed in first basic part ("generalized Peres-Horodecki conditions") of the paper, in which we use convex combinations of quantum-mechanical terms
to form nonnegativity constraints.

\begin{figure}
\includegraphics{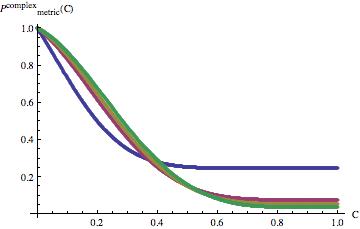}
\caption{\label{fig:bothcomplex}Separability probabilities of generic complex
two-qubit states having concurrence less than or equal to $C$. 
The Hilbert-Schmidt (blue) curve intersects the other three. 21,100,000 
TF-points were employed.}
\end{figure}

\begin{figure}
\includegraphics{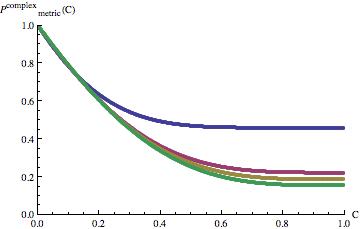}
\caption{\label{fig:bothreal}Separability probabilities of generic real
two-qubit states having concurrence less than or equal to $C$. The Hilbert-Schmidt (blue) curve intersects the other three 
from below, all near $\alpha=0.12$. There were 7,800,000 
TF-points employed.}
\end{figure}

\begin{figure}
\includegraphics{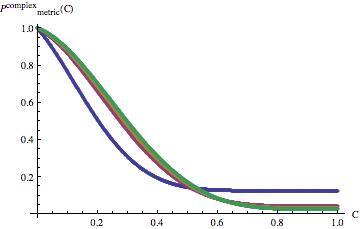}
\caption{\label{fig:bothcomplexarea}Separability probabilities of generic 
rank-3 complex
two-qubit states having concurrence less than or equal to $C$. The Hilbert-Schmidt (blue) curve intersects the other three. 14,000,000 TF-points were employed.}
\end{figure}

\begin{figure}
\includegraphics{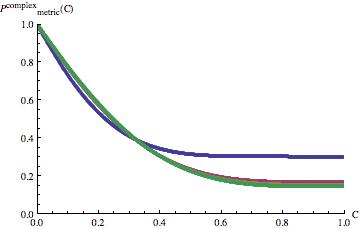}
\caption{\label{fig:bothrealarea}Separability probabilities of generic 
rank-3 real
two-qubit states having concurrence less than or equal to $C$. The Hilbert-Schmidt (blue) curve intersects the other three. 
9,750,000 TF-points were employed.}
\end{figure}

Now, we simply estimate--again, with respect to the four metrics
in question--the separability probability of two-qubit states
for which the concurrence $C$ is less than some threshold $C_{0}$.
We show our results in Figs.~\ref{fig:bothcomplex} and \ref{fig:bothreal}.
The generic rank-3 counterparts of these two figures are given in 
Figs.~\ref{fig:bothcomplexarea} and \ref{fig:bothrealarea}.
In all four of these cases, the Hilbert-Schmidt curve intersects the curves for the three monotone metrics from below. 
In this regard, it has been 
noted by Bengtsson and {\.Z}yczkowski that the "Bures measure is 
concentrated at the states of higher [than the Hilbert-Schmidt] 
purity"
\cite[p. 356]{ingemarkarol}, since $\langle \mbox{Tr} \rho^2 \rangle_{HS}  
<  \langle \mbox{Tr} \rho^2 \rangle_{Bures}$. Our (intersecting) results in this 
set of concurrence-based analyses is clearly consistent--but now 
taking a separability-related form--with that 
assertion.
\section{Remarks} \label{Remarks}
Our motivation in undertaking the first principal part of this study 
reported above has been to examine whether it might be feasible
to shift the question of 
determining the two-qubit separability probabilities with respect
to various metrics of quantum-mechanical interest 
to the (perhaps more tangible, addressable) 
question of characterizing the {\it curves} that {\it interpolate}
between such separability probabilities and the (unit) probabilities that a
two-qubit state is {\it either} separable {\it or} entangled (cf. \cite{szHS,szBures}).
We intend to study the curves generated in still greater detail, as additional 
computations render them more precise. (The Tezuka-Faure procedure 
is not amenable to use of {\it statistical} tests, though variants of this
quasi-Monte Carlo method have been developed that are.) In particular, it would be of interest
to see if the differences between the curves for the three monotone metrics 
studied could be explained directly in terms of the Chentsov-Morozova functions
$c(x,y)$
for those metrics \cite{petzsudar,lesniewski}. These are $\frac{2}{x+y}$, 
$\frac{4}{(\sqrt{x} +\sqrt{y})^2}$ and $\frac{(\log{x}-\log{y})}{x-y}$, for
the Bures, Wigner-Yanase and Kubo-Mori metrics, respectively. (The associated 
operator monotone functions, $f(t)$, for which $c(x,y) =\frac{1}{y f(\frac{x}{y})}$
are $\frac{1+t}{2}$, $\frac{t+2 \sqrt{t}+1}{4}$ and $\frac{t-1}{\log{t}}$, respectively.) The possible relevance of the Dyson-index-{\it  ansatz} to explain
differences between results for the generic real and generic complex
systems--as in sec.~\ref{secpiecewise} below--should also be 
examined \cite{slater833,maxconcur4}.

Conceiveably, our attempted generalizations here of the Peres-Horodecki 
conditions and introduction of the concept of "$\alpha$-separability" might prove productive
in some manner parallel to the well-studied 
concepts (also based on generalizations/extensions/embeddings) 
of $p$-R{\'e}nyi-entropy 
\cite{Alicki} and of escort distributions \cite{escort,slaterHusimi}.

\section{Separabilities as piecewise continuous functions of maximal concurrence} 
\label{secpiecewise}
\subsection{Objective}
Here, we begin the second basic part of our paper. We importantly amend a certain parenthetical remark made
in our recent paper \cite{maxconcur4}, to the effect 
that although two-qubit {\it diagonal-entry}-parameterized separability
functions (DESFs) had been shown 
\cite{slater833,slaterJGP2} to clearly conform to a pattern dictated by 
the ``Dyson indices'' ($\beta = 1$ [real], 2 [complex],
4 [quaternionic])
of random matrix theory, 
this did not appear to be the case with 
regard to {\it eigenvalue}-parameterized separability functions (ESFs). (We remark here that the 
"value of $\beta$ is given by the number of independent
degrees of freedom per matrix element and is determined by the antiunitary 
symmetries \ldots It is a concept that originated in Random Matrix Theory 
and is important for the Cartan classification of symmetric spaces'' 
\cite[p. 480]{kogut}. The Dyson index corresponds to the ``multiplicity of 
ordinary roots'', in the terminology of symmetric spaces \cite[Table 2]{caselle}.) 
But upon further examination of the extensive numerical analyses 
reported in \cite{maxconcur4}, 
we found quite convincing evidence that adherence to the Dyson-index 
pattern does also hold for ESFs,
at least as regards the 
upper {\it half}-range 
$\frac{1}{2} \leq C_{max} \leq 1$ of the maximal concurrence 
over spectral orbits (\ref{maximalconcurrenceformula}).

To be specific, it strongly appears 
that in this upper half-range, the {\it real} two-qubit ESF is 
equal to 
to $\frac{(2-2 C_{max})^{\frac{3}{2}}}{\sqrt{30}}$, and its {\it complex} counterpart--in 
conformity to the Dyson-index pattern--proportional
to the {\it square} of the real ESF, 
that is, $\frac{(2-2 C_{max})^3}{15}$. The previously documented piecewise 
continuous (``semilinear'') behavior
in the {\it lower} half-range $0 \leq C_{max} \leq \frac{1}{2}$ appeared 
to lack any particular Dyson-index-related interpretation--which seemed
somewhat paradoxical in terms of our DESF-findings 
\cite{slater833,slaterJGP2}. However, we report new insights into this problem below 
(sec.~\ref{lowerpart}).
\subsection{Previous ESF findings}
The study \cite{maxconcur4} had been devoted to the
question of determining for the generic 
(9-dimensional) real and (15-dimensional)
complex two-qubit systems, the nature of certain 
trivariate ``eigenvalue-parameterized separability
functions'' (ESFs). These 
(metric-{\it independent}) 
ESFs, it was argued, could substantially 
assist in the determination of
separability 
{\it probabilities} in terms of certain metrics (the Hilbert-Schmidt
and Bures being the most conspicuous examples). 
(In \cite{slater833}, DESFs were successfully used in the Hilbert-Schmidt
case, but they do not seem as useful for the Bures and other montone metrics, the standard formulas for which are expressed in terms of eigenvalues, and {\it not} diagonal entries.)
We further investigated 
in \cite{maxconcur4} the possibility that these {\it prima facie} 
trivariate functions of the eigenvalues $\lambda_i$ $(i =1,\ldots 4)$ 
of $4 \times 4$ density matrices $(\lambda_4=1-\Sigma_i^3 
\lambda_i)$, were expressible as {\it univariate} 
functions 
\begin{equation} \label{ansatz}
S_4^{(\beta)}(\lambda_1\ldots\lambda_4) = \sigma^{(\beta)}
(C_{max}(\lambda_1\ldots\lambda_4)),
\end{equation}
of the {\it maximal concurrence} $C_{max}$--given by 
(\ref{maximalconcurrenceformula})--over
spectral orbits  \cite[sec. VII]{roland2} \cite{ishi,ver}. 
(At this point in our presentation, let us--motivated 
by Dyson-index conventions--regard 
$\beta$ in (\ref{ansatz}) 
only as a notational [dummy variable], not calculational device taking
the values 1 [real], 2 [complex], 4 [quaternionic].)
\subsection{Jump discontinuities}
Our main conclusions in \cite{maxconcur4} 
were that--if the reducibility-to-univariance property 
(\ref{ansatz}) held, as our extensive numerical evidence 
appeared to suggest might be the case (being able to explain almost $99\%$ of the variance \cite[Sec. II.B.1]{maxconcur4})--the 
associated real and complex 
univariate functions both had jumps of approximately
$50\%$ magnitude at $C_{max}=\frac{1}{2}$, as well as a number of additional
discontinuities (remarkably 
coincident in both the real and complex cases) in the 
{\it lower} half-range $C_{max} \in (0,\frac{1}{2}]$. 
(The joint jumps at $C_{max}=\frac{1}{2}$ were displayed in \cite{maxconcur4} in
Figs. 2 and 6. We have since found a small programming error
that caused the two curves in Fig. 2 there to be slightly 
more misaligned--by $\frac{1}{500}$--than they should have been.) 
Also, both univariate functions
appeared to be 
simply {\it linear} between certain of these discontinuities. 
The {\it upper} half-range $C_{max}
\in [\frac{1}{2},1]$--in which the univariate functions 
of $C_{max}$ took lesser values--did 
not command our attention 
in \cite{maxconcur4}, seeming to be of relatively less 
interest. Our only pertinent observation there was that
there did not appear to be any discontinuities 
in that segment.
\subsection{New Dyson-index-related findings} \label{New}
\begin{figure}
\includegraphics{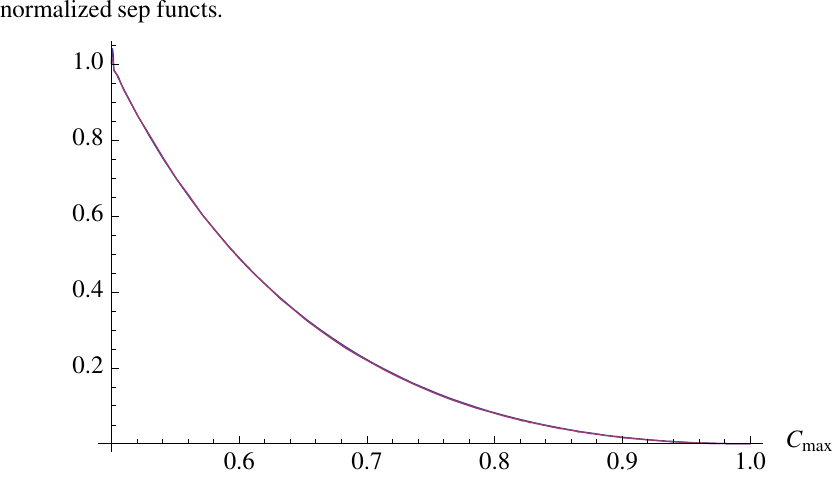}
\caption{{\it Joint} plot of numerical estimates of
$\Big( \frac{\sigma^{1}(C_{max})}{\sigma^{1}(\frac{1}{2})} \Big)^2$ {\it and}
$ \frac{\sigma^{2}(C_{max})}{\sigma^{2}(\frac{1}{2})}$ for 
$C_{max} \in [\frac{1}{2},1]$. 
\label{fig:normed}}
\end{figure}

\begin{figure}
\includegraphics{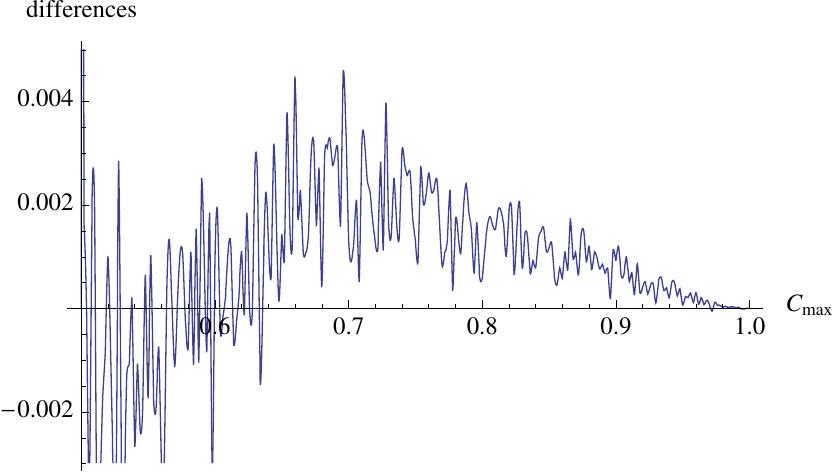}
\caption{\label{fig:newDyson}Numerical estimate of 
$\Big( \frac{\sigma^{1}(C_{max})}{\sigma^{1}(\frac{1}{2})} \Big)^2 -
 \frac{\sigma^{2}(C_{max})}{\sigma^{2}(\frac{1}{2})}$}
\end{figure}

\begin{figure}
\includegraphics{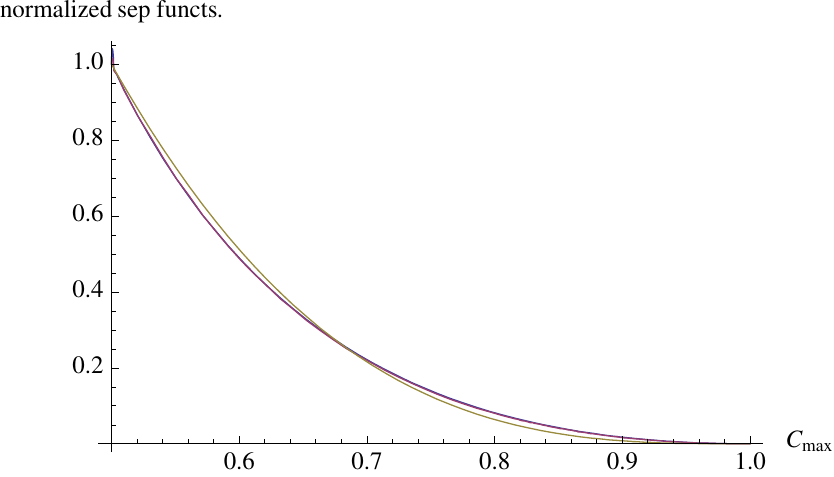}
\caption{\label{fig:normed2}The two functions in Fig.~\ref{fig:normed}, 
along with the 
additional ({\it very} 
closely-fitting) function $(2 -2 C_{max})^3$.}
\end{figure}

Now, in fact, 
turning our attention 
more closely to this upper half-range 
$C_{max} \in [\frac{1}{2},1]$, we readily find
strong evidence for a very interesting Dyson-index-type phenomenon.
If we {\it normalize} our extensive 
numerical estimates from \cite{maxconcur4} 
of $\sigma^{1}(C_{max})$ and  $\sigma^{2}(C_{max})$
to both equal 1 at the {\it jump} discontinuity point 
$C_{max}=\frac{1}{2}$, then a joint plot 
(Fig.~\ref{fig:normed}) of the latter 
normalized (complex) function versus the {\it square} of the former 
normalized 
(real) function for $C_{max} \in [\frac{1}{2},1]$ remarkably 
shows {\it no} perceptible difference between the two resulting curves. 
(The sample 
[quasi-Monte Carlo] estimate of 
$\sigma^{2}(\frac{1}{2})$ is 0.0651586 and that of 
$\sigma^{1}(\frac{1}{2}))$ is 0.1803748.)
In Fig.~\ref{fig:newDyson},
we show--on a much finer scale than used 
in Fig.~\ref{fig:normed}--the actual (very small) 
numerically-obtained differences 
\begin{equation}
\Big( \frac{\sigma^{1}(C_{max})}{\sigma^{1}(\frac{1}{2})} \Big)^2 -
 \frac{\sigma^{2}(C_{max})}{\sigma^{2}(\frac{1}{2})}
\end{equation}
between them.
Of further 
considerable importance, Fig.~\ref{fig:normed2} is a repetition of 
Fig.~\ref{fig:normed}, but along with the insertion now 
of the function
\begin{equation} \label{simple}
(2 -2 C_{max})^3 = 8 (1-C_{max})^3,
\end{equation}
which we see fits our two estimates {\it very} well.
Assuming that (\ref{simple}) is the correct form 
(up to the still 
not exactly-known normalization factor) of $\sigma^2(C_{max})$ over 
$C_{max}\in [\frac{1}{2},1]$, we can estimate the 
associated contribution to the separability probabilities from density matrices corresponding to 
this half-range 
to the Hilbert-Schmidt and Bures separability probabilities of
generic complex two-qubit systems to be 0.0100578 and 0.0194829, respectively.
(The real counterparts 
of these separability probabilities are, then, 
0.0254346 and 0.0100578, respectively [cf. (\ref{Ereal}), (\ref{Ecomplex})].)

Let us further note that our sample estimate of
the ratio
\begin{equation} \label{ratio}
 \frac{\sigma^{2}(\frac{1}{2})}{\Big(\sigma^{1}(\frac{1}{2})\Big)^2}
= \frac{0.0651586}{0.1803748^2} = 2.00272
\end{equation}
is very close (and possibly theoretically exactly 
equal) to 2.

Over $0 \leq C_{max} \leq \frac{1}{2}$, the range of primary 
interest in \cite{maxconcur4}, the estimates of
the real and complex two-qubit separability functions {\it intersect} 
(near $C_{max}=0.1812$),
and appear to have linear segments 
over the {\it same} subintervals \cite[Figs. 1, 5, 7]{maxconcur4}. 
These features appeared to make any immediate application of
the Dyson-index pattern problematical in this lower half-range.
So, the behaviors of the univariate functions $\sigma^{(\beta)}(C_{max})$, 
($\beta =1$ [real], 2 [complex]),
over the two indicated regimes of $C_{max}$ seem to be highly distinct.
The point $C_{max}=\frac{1}{2}$ clearly serves as a point of 
major behavioral transition, 
with the lower half-range, then, appearing 
perhaps to be the more theoretically
challenging of the two. (We will observe what appears to
be similarly dichotomous Dyson-index behavior in the qubit-qutrit
case [Fig.~\ref{fig:maxconcurRank6}]. Perhaps one might view the two 
regimes as semiclassical and quantum in nature.)

An outstanding
question is what are the specific values of 
$\sigma^{1}(\frac{1}{2})$ and $\sigma^{2}(\frac{1}{2})$, which 
we used as normalization factors in our analyses above. The
nearness to 2 of the ratio (\ref{ratio}) may be a helpful guide 
in this regard. In fact,  
let us take this opportunity to further indicate that in our
ongoing supplemental analyses--in which we use 5,000, rather than
500 sampling points in 
the interval [0,1]--we have obtained for the ratio (\ref{ratio})
the estimate
\begin{equation} \label{ratio2}
\frac{\sigma^{2}(\frac{1}{2})}{\Big(\sigma^{1}(\frac{1}{2})\Big)^2}
= \frac{0.066663}{0.18275^2} = 1.99605.
\end{equation}
This ratio would be exactly 2 if we took for the numerator of (\ref{ratio2}) the value 
$\frac{1}{15} \approx 0.0666667$ and for the denominator, 
$\frac{1}{30} \approx 0.182574^2$. We will, in fact, assume these exact values 
in seeking to ascertain in sec.~\ref{newresults}, the exact contributions over $C_{max} \in [\frac{1}{2},1]$ to the total Hilbert-Schmidt two-qubit generic real 
and complex separability probabilities.

\subsection{Rank-three complex and real two-qubit cases} \label{rankthree}
\begin{figure}
\includegraphics{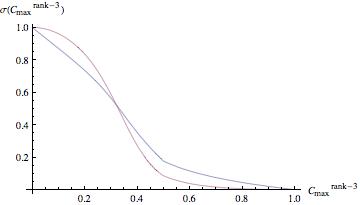}
\caption{\label{fig:maxconcurRank3}Eigenvalue-parameterized separability
functions, expressed in terms of maximal concurrence, for the {\it minimally degenerate rank-three} generic real and 
complex two-qubit states. The complex (red) curve is initially
dominant. For each TF-point employed (2,062,400 in the complex case and 2,331,300 in the real case), separability is checked for 500 
equally-spaced values of $C_{max}^{rank-3}$. There appear to be 
discontinuities at $C_{max}^{rank-3} = \frac{1}{2}$.}
\end{figure}

Here, we straightforwardly apply the same 
maximal-concurrence ansatz (\ref{ansatz})  just discussed 
and applied to 
the full (rank-4) complex and real two-qubit cases,
to the minimally degenerate (rank-3) counterparts.
The main conceptual point to note is that the formula
for the maximal concurrence (\ref{maximalconcurrenceformula}) 
now degenerates to 
\begin{equation} \label{maximalconcurrenceformularank3}
C_{max}^{rank-3}=\lambda_1-\lambda_3, \hspace{.5in}
(\lambda_1 \geq \lambda_2 \geq \lambda_3 ).
\end{equation}
In Fig.~\ref{fig:maxconcurRank3} we show
the joint plot of the corresponding real and complex curves. The 
estimated complex (red) curve initially dominates the estimated 
real (blue) curve 
(cf. \cite[Fig. 1]{maxconcur4}).
\subsubsection{Close resemblance to generic rank-4 Dyson-index pattern} \label{rankthreesub}
\begin{figure}
\includegraphics{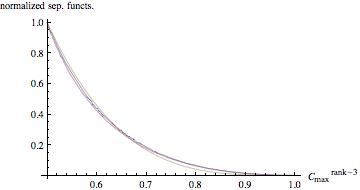}
\caption{\label{fig:Rank3counterpart}The rank-3 counterpart to  
Fig.~\ref{fig:normed2}, with the rather well-fitting function 
$(2- 2 C_{max}^{rank-3})^\frac{7}{2}$ 
replacing $(2- 2 C_{max}^{rank-4})^3$.}
\end{figure}

It appears now--as a plot (Fig.~\ref{fig:Rank3counterpart}) parallel to that displayed in Fig.~\ref{fig:normed2} indicates--that the Dyson-index 
pattern continues to hold for the range $C_{max}^{rank-3} \in 
[\frac{1}{2},1]$ in the two-qubit generic rank-3 cases, but with the replacement of 
$(2- 2 C_{max}^{rank-4})^3$ by  $(2- 2 C_{max}^{rank-3})^\frac{7}{2}$.

\begin{figure}
\includegraphics{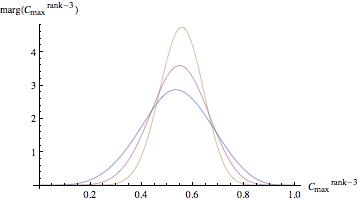}
\caption{\label{fig:marginalConc}Univariate marginal probability distributions (\ref{conctransf1}) and (\ref{conctransf2}) over $C_{max}^{rank-3}$ of the Hilbert-Schmidt measures on the eigenvalues. 
The quaternionic curve has the highest peak and the real curve, the 
lowest. All three curves are asymmetric about $C_{max}^{rank-3}=\frac{1}{2}$ and skewed to the right. The real peak (mode) is at 0.534989, the complex at 0.549857, and the quaternionic at 0.558738, while the medians 
are 0.53707, 0.5483 and 0.556787, respectively. Also, the means 
are $\frac{781}{2 \cdot 3^6} \approx 0.535665, \frac{35}{2^6} \approx 0.546875$, and  
$\frac{27313}{2^{14} \cdot 3} \approx 0.555684$, respectively.}
\end{figure}

To test the possible applicability of the rank-3 version of the univariance hypothesis (\ref{ansatz}), we estimated
the real and complex rank-3 two-qubit Hilbert-Schmidt separability probabilities using
the ESFs displayed in Fig.~\ref{fig:maxconcurRank3}. The values we 
obtained were 0.208172 and 0.104852, respectively (while the correponding
conjectured values were perhaps somewhat disappointingly different, calling for further analysis, that is
$\frac{4}{17} \approx 0.235294$ and $\frac{4}{33} \approx 0.121212$).
We can express these results as {\it one}-dimensional integrals  
over $C_{max}^{rank-3} \in [0,1]$ of the product of the real function displayed in 
Fig.~\ref{fig:maxconcurRank3}, and (using $C \equiv C_{max}^{rank-3}$) 
the univariate marginal Hilbert-Schmidt 
probability distribution (Fig.~\ref{fig:marginalConc})
\begin{equation} \label{conctransf1} 
\mbox{marg}_{real}(C)=
\begin{cases}
 -\frac{1792}{81} C^4 \left(12 C^2-5\right) & 0<C\leq \frac{1}{2} \\
 \frac{3584}{81} (C-1)^4 C (4 C (5 C-1)-1) & \frac{1}{2}<C<1
\end{cases}
\end{equation}
and the integral over $C_{max}^{rank-3} \in [0,1]$ of the product of the complex function displayed in 
Fig.~\ref{fig:maxconcurRank3}, and the univariate marginal 
Hilbert-Schmidt probability distribution 
(Fig.~\ref{fig:marginalConc})
\begin{equation} \label{conctransf2}
\mbox{marg}_{complex}(C)=
\begin{cases}
 -\frac{7280}{729} (C-1)^7 C^2 (C (C (C (16325 C-7693)-379)+315)+45) &
   \frac{1}{2}<C<1 \\
 \frac{7280}{729} C^7 \left(155 C^6+1287 C^4-1089 C^2+231\right) &
   0<C\leq \frac{1}{2}
\end{cases}.
\end{equation}
(These distributions are "marginal", in the sense that they are obtained by integrating the  HS or Bures measure defined on the three-dimensional simplex of eigenvalues--obtainable from the papers of {\.Z}yczkowski and Sommers \cite{szHS,szBures}--over two of the three coordinates [the third coordinate being $C_{max}$] used to parameterize the simplex.)
Also, we have for $0<C<\frac{1}{2}$, 
\begin{equation} \label{conctransf3}
\mbox{marg}_{quaternionic}(C)=
\end{equation}
\begin{displaymath}
\frac{9209200 C^{13} \left(3 \left(7133 C^{10}+236790 C^8+253023
   C^6-729980 C^4+497097 C^2-142766\right) C^2+46189\right)}{531441}.
\end{displaymath}
(The quaternionic expression for $\frac{1}{2}<C<1$ is somewhat more cumbersome in
nature to present.)
To obtain these univariate functions, we have transformed one of
the eigenvalues, say $\lambda_1$ to $C_{max}^{rank-3} 
\equiv \lambda_1- \lambda_3$ (the jacobian of the transformation being 
unity) and integrated (restricted to the Weyl chamber of ordered 
eigenvalues) the corresponding (bivariate in this case) 
Hilbert-Schmidt measures (over the eigenvalues) \cite[eqs. (4.1), (6.5), (7.8)]{szHS} over 
$\lambda_2$. Fitting the means and variances of ((\ref{conctransf1})-(\ref{conctransf3})), we can obtain beta distribution $B(p,q)$ approximations to
the real, complex and quaternionic probability distributions using the paired sets of parameters 
$\{p,q\}= \left\{\frac{47641}{7196},\frac{41297}{7196}\right\} \approx 
 \left\{6.62048, 5.73888 \right\}, \left\{\frac{12323885}{1142816},\frac{10211219}{1142816}\right\} \approx 
 \left\{10.7838,8.93514 \right\}$ and 
$\left\{\frac{4108424031600889}{214515575216232},\frac{3285024436207367}{2
   14515575216232}\right\} \approx \left\{19.1521, 15.3137 \right\}$, respectively. 
Beta distributions, defined over the unit interval, are a general type of statistical distribution, related to the gamma distribution, and have two free parameters.

\begin{figure}
\includegraphics{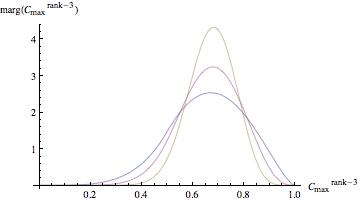}
\caption{\label{fig:marginalConcBures}Counterpart to 
Fig.~\ref{fig:marginalConc} based on the Bures metric. The order of curves is the same. Numerical methods were employed.}
\end{figure}

The Hilbert-Schmidt (total--separable and nonseparable) probability that a minimally-degenerate two-qubit
state has maximal concurrence within the range $[\frac{1}{2},1]$ is 
$\frac{49}{81} =(\frac{7}{9})^2 \approx 0.604938$ for real states, 
$\frac{996431}{2^{11} \cdot 3^6} \approx 0.667405$ for complex states, 
and $\frac{3335170241153}{2^{23} \cdot 3^{12}} \approx 0.748123$ 
for quaternionic states. (For the [smaller] full-rank counterparts
see sec.~\ref{Total}.)
In Fig.~\ref{fig:marginalConcBures}, we show the counterpart to Fig.~\ref{fig:marginalConc} based on the Bures (minimal monotone) metric.

Let us assume (cf. Fig.~\ref{fig:Rank3counterpart}) 
that the ESF in the real case is proportional over
$C_{max}^{rank-3} \in [\frac{1}{2},1]$ to $(2 -2 C_{max}^{rank-3})^\frac{7}{4}$  and in the complex case to the square of this. Then, we have
that the contributions over this half-domain to the Hilbert-Schmidt real and complex separability
probabilities, respectively, are $\frac{2^6 \cdot 7 \cdot 1249}{3^6 \cdot 5 \cdot 13 \cdot 31} \approx 0.380924$ (multiplied by a normalization constant approximately   
 0.177365) and $\frac{13 \cdot 289014610051}{2^9 \cdot 3^9 \cdot 5 \cdot 11 \cdot 23 \cdot 29 \cdot 31} \approx 0.327832$ (multiplied by a normalization constant approximately 0.086232).
\subsection{Rank-five complex qubit-qutrit case} \label{rankfive}
For the full-rank qubit-qutrit case, the counterpart--although 
not enjoying all the properties--of 
the two-qubit maximal concurrence formula 
(\ref{maximalconcurrenceformula}) is \cite[p. 16]{roland2}
\begin{equation} \label{maximalconcurrenceformulaRank6}
C_{max}^{rank-6}=\mbox{max}(0,\lambda_1-\lambda_5 
-2 \sqrt{\lambda_4 \lambda_6}), \hspace{.5in}
(\lambda_1 \geq \lambda_2 \geq \lambda_3 \geq \lambda_4 \geq \lambda_5 \geq \lambda_6),
\end{equation}
which, obviously (since $\lambda_6=0$), degenerates (using the same 
eigenvalue-ordering) to
\begin{equation} \label{degenerate}
C_{max}^{rank-5}=\lambda_1-\lambda_5.
\end{equation}

\begin{figure}
\includegraphics{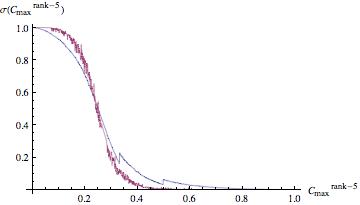}
\caption{\label{fig:maxconcurRank5} Qubit-qutrit 
eigenvalue-parameterized separability 
functions as a function of maximal concurrence for
the generic rank-5 real and complex $6 \times 6$ density matrices. 104,200
TF-points were used in the real case and 181,600 in the 
complex case, while, for each such 
point, the interval [0,1] was 
sampled at 1,000 equally-spaced locations of $C_{max}^{rank-5}$. 
The complex (red) curve is initially dominant. There are manifest 
discontinuities (jumps) at $C_{max}^{rank-5} = \frac{1}{3}$ and 
$\frac{1}{2}$ in the real case.}
\end{figure}

In Fig.~\ref{fig:maxconcurRank5}--again under the hypothesis (ansatz)
that the corresponding eigenvalue-parameterized-separability 
function is a (univariate) function 
of the maximal concurrence 
expression (\ref{degenerate})--we 
show the
analogue of Fig.~\ref{fig:maxconcurRank3} 
for the minimally-degenerate (rank-5)
generic real and complex qubit-qutrit case. 
(Again, in the complex case we used the $SU(6)$-Euler-angle 
parameterization of S. Cacciatori \cite[App. A]{JMP2008}, 
while we used a yet unpublished
Euler-angle parameterization of his of $SO(6)$ for the real case.)
There are evident jumps in the real (blue) curve 
at $C_{max}^{rank-5} = \frac{1}{3}$ and
$\frac{1}{2}$. 
(The still erratic nature of the complex [red] curve--we used 1,000 [not 500] equally-spaced points in [0,1]--makes it, at this point of sampling, difficult to gauge the applicability of the Dyson indices.)
To test the applicability of the rank-5 version of the univariance hypothesis (\ref{ansatz}), we estimated
the real and complex rank-5 qubit-qutrit Hilbert-Schmidt separability probabilities using
the ESFs displayed in Fig.~\ref{fig:maxconcurRank5}. The values we 
obtained were 0.097232 and 0.0226654, respectively, while the corresponding
conjectures were--again somewhat disappointingly different--$\frac{16}{213} \approx 0.0751174$ and $\frac{16}{1199} \approx 0.0133445$.

\subsection{Full-rank real and complex qubit-qutrit cases} \label{ranksix}
In the two-qubit case, we had evolved a computational strategy
in which we used the Mathematica command FindInstance to 
systematically generate 
random sets of  three or four eigenvalues that yielded values of 
the maximal concurrence ((\ref{maximalconcurrenceformula}) or 
(\ref{maximalconcurrenceformularank3})) 
at {\it equally-spaced} intervals $C_{max} \in [0,1]$--and, similarly, 
in the rank-5 qubit-qutrit case (\ref{degenerate}).
However, due to greater complexity in the rank-6 case, this strategy did not prove at all
feasible for generating random sets of {\it six} eigenvalues yielding
equally-spaced values of the maximal 
concurrence (\ref{maximalconcurrenceformulaRank6}).

\begin{figure}
\includegraphics{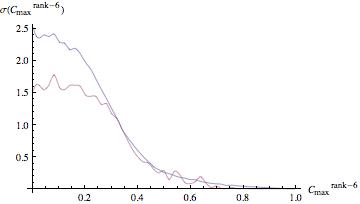}
\caption{\label{fig:maxconcurRank6} Joint plot 
(cf. Fig.~\ref{fig:normed})--to test Dyson-index ansatz--of numerical estimates of the real 
$\Big( \frac{\sigma^{1}(C_{max}^{rank-6})}{\sigma^{1}(\frac{1}{3})} \Big)^2$  and complex
$\frac{\sigma^{2}(C_{max}^{rank-6})}{\sigma^{2}(\frac{1}{3})}$ .  (The complex [red] curve is lower at $C_{max}^{rank-6}=0$.) There were 62,086,051    
20-dimensional TF-points (each point corresponding to a single 
density matrix) used in the real case and 451,373,489 
35-dimensional TF-points in the 
complex case. Each of these points 
was allocated to one of fifty bins in the 
interval $C_{max}^{rank-6 } \in [0,1]$.}
\end{figure}

\begin{figure}
\includegraphics{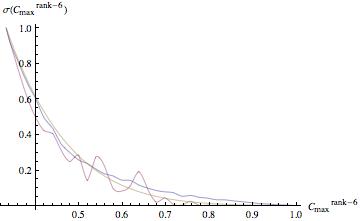}
\caption{\label{fig:QubQutAll}Same plot as Fig.~\ref{fig:maxconcurRank6}, but restricted to $C_{max}^{rank-6} \in[\frac{1}{3},1]$, along with the insertion of the smooth, interweaving function
 $(\frac{4}{3} - C_{max}^{rank-6})^7$}.
\end{figure}

So, we altered our approach, now simply 
randomly generating density matrices (again using the same quasi-Monte Carlo routines \cite{giray1}) and 
recording their associated values of concurrence. We "binned" these 
concurrence 
values into intervals of length $\frac{1}{50}$, and averaged the total
measures recorded by the number of observations within the individual bins.   We interpolated these average values to obtain the 
associated eigenvalue-parameterized separability functions (ESFs).
We have generated the corresponding curves for both the full-rank
real and complex generic qubit-qutrit cases, but they are still somewhat
crude/rough in character. 
Nevertheless, we plot in Fig.~\ref{fig:maxconcurRank6} 
(cf. Figs.~\ref{fig:normed} and \ref{fig:Rank3counterpart}) normalized forms of the complex (red)
curve and the {\it square} of the real (blue) curve. They appear to indicate possible adherence to the Dyson-index ansatz, since the two curves
closely "track" each other, at least (as our generally observed  pattern in the two-qubit case would suggest) for the {\it higher} values of $C_{max}^{rank-6}$. (It seems
that this domain of possibly strict Dyson-index behavior may be $C_{max}^{rank-6} \in 
[\frac{1}{3},1]$, while in the full-rank two-qubit case 
(Fig.~\ref{fig:normed}) it highly convincingly appeared to be $C_{max}^{rank-4} \in [\frac{1}{2},1]$. Our level of binning is perhaps too coarse for 
the detection of possible discontinuities in the two curves.)
We were interested in seeing how close the plotted curves came--under the rank-6 qubit-qutrit version of the univariance hypothesis (\ref{ansatz})--to yielding the conjectured HS real and complex separability probabilities of $\frac{32}{213}$ and $\frac{32}{1199}$ \cite[sec. X]{slater833}, but the requisite numerical integrations proved quite problematical to perform.
In Fig.~\ref{fig:QubQutAll} we plot the two functions in 
Fig.~\ref{fig:maxconcurRank6} over the interval $[\frac{1}{3},1]$, along
with the interweaving curve $(\frac{4}{3} -C_{max}^{rank-6})^7$.
\section{Separability probability decompositions over $C_{max}$ regions} \label{newresults}
\subsection{$C_{max}=0$ domain} \label{CmaxDomain}
One can--in an apparently natural manner--consider the two-qubit real, complex 
and quaternionic Hilbert-Schmidt 
separability probabilities to be the sum of {\it three} components: (1)
the Hilbert-Schmidt {\it absolute} 
separability probabilities (corresponding
to $C_{max}=0$); (2) the probabilities over the range $C_{max} \in (0,\frac{1}{2}]$; and (3) the probabilities over the range $C_{max} \in [\frac{1}{2},1]$. (For a contour plot of the three-dimensional body $C_{max}=0$, 
see \cite[Fig. 2]{JMP2008}.) 
Now, we have previously been able to compute the absolutely separable components 
\cite[eqs. (34), (35)]{JMP2008}.
These are 
\begin{equation} \label{HSlower}
P^{HS_{real}}_{C_{max}=0}= \frac{6928-2205 \pi }{2^{\frac{9}{2}}} 
\approx 0.0348338,
\end{equation}
\begin{equation} \label{HSupper}
P^{HS_{complex}}_{C_{max}=0} = 
\frac{\psi_1- \psi_2 \sqrt{2}- \psi_3 \sqrt{2} \pi
   + \psi_4 \sqrt{2} \sec ^{-1}(3)}{2^{16} 3^5} \approx 
0.0036582630543035
\end{equation}
(the Bures [minimal monotone] metric analogue being 
considerably smaller, 0.000161792 \cite[p. 25]{JMP2008})
where
\begin{equation*}
\psi_1=956877309536,  \hspace{.1in} \psi_2=781862943168,
 \end{equation*}
\begin{equation*}
\psi_3=746624752335, \hspace{.1in} \psi_4=1990999339560,
\end{equation*}
and 
\begin{equation} \label{HSmost}
P^{HS_{quat}}_{C_{max}=0} = 
-\frac{13 \left(\phi_1+\phi_2
   \sqrt{2}+ \phi_3 \sqrt{2} \pi
   -\phi_4 \sqrt{2} \sec
   ^{-1}(3)\right)}{2^{34} \cdot 3^{11}}
\approx 0.0000398703,
\end{equation}
where 
\begin{equation}
\phi_1=-806338156306739134839776, \hspace{.1in}  \phi_2=
658857590468226345222144, 
\end{equation}
\begin{displaymath}
\phi_3=629162653900414735065195, \hspace{.1in} \phi_4=
1677767077067772626840520.
\end{displaymath}
(These are "conjecture-free" results, not dependent on any Dyson-index 
ansatz. In \cite[eqs. (36), (37)]{JMP2008} 
we gave a considerably lengthier, but fully equivalent, expression for 
$P^{HS_{quat}}_{C_{max}=0}$. One might seek to find explanations for the large integers 
displayed above in terms of gamma functions. The computational challenges to computing analogous 
absolute separability results
for the qubit-qutrit states appear to be highly formidable.)
\subsection{$C_{max} \in [\frac{1}{2},1]$}
Further, accepting the strongly-supported Dyson-index ansatz 
(Fig.~\ref{fig:normed2} and (\ref{ratio2})) that 
$\sigma^{(1)}(C_{max}(\lambda_1\ldots\lambda_4))
 = \frac{(2-2 C_{max})^{\frac{3}{2}}}{\sqrt{30}}$  and 
$\sigma^{(2)}(C_{max}(\lambda_1\ldots\lambda_4))= \frac{(2-2 C_{max})^3}{15}$ for $C_{max} \in [\frac{1}{2},1]$, we can now add to the absolute separability probabilities ($C_{max}=0$) listed immediately above, 
the conjectured probability contributions
\begin{equation} \label{Ereal}
P^{HS_{real}}_{C_{max} \in [\frac{1}{2},1]}= \frac{\sqrt{\frac{3}{10}} \left(3162214-738885 \sqrt{2} \tan
   ^{-1}\left(\sqrt{2}\right)\right)}{2^{12} \cdot 5 \cdot 7 \cdot 17 \cdot 19} \approx 
0.02559647778,
\end{equation}
and
\begin{equation} \label{Ecomplex}
P^{HS_{complex}}_{C_{max} \in [\frac{1}{2},1]}= \frac{7 \left(148453588142-79729806357 \sqrt{2} \cot
   ^{-1}\left(\frac{5}{\sqrt{2}}\right)\right)}{2^{31} \cdot 3^7 \cdot 17} 
\approx 0.01029059519.
\end{equation}
Further, using the Dyson-index ansatz with $\beta=4$, we obtain 
\begin{equation} \label{Equat}
P^{HS_{quat}}_{C_{max} \in [\frac{1}{2},1]}=  
\frac{5 \left(\zeta_1-\zeta_2
   \sqrt{2} \cot
   ^{-1}\left(\frac{5}{\sqrt{2}}\right)\right)}{2^{66} \cdot 3^8 \cdot 11 \cdot 29 \cdot 31} \kappa \approx 0.165191  \kappa,
\end{equation}
where 
\begin{equation} \label{Equat2}
\zeta_1 = 174916374035295022487516506, \hspace{.1in} 
\zeta_2 = 42964561240209557008032951,
\end{equation}
and $\kappa$ is the $\beta=4$ unknown and 
yet-unconjectured analogue 
of the presumed real and complex constants $\frac{1}{\sqrt{30}}$ and $\frac{1}{15}$.
(In computing (\ref{Ereal})-(\ref{Equat2}), we found 
a joint transformation of the form 
$\alpha_1=\sqrt{\frac{\lambda_2}{\lambda_4}}$ and 
$\alpha_2=\sqrt{\lambda_2 \lambda_4}$ to be helpful.) 
\subsubsection{Corollaries to the "twofold" SBZ-Theorem}
Since the probability is zero that a generic minimally-degenerate two-qubit state
is absolutely separable (that is, $P^{rank-3} =P^{rank-3}_{C_{max} \in [0,1]}$)--as can be immediately deduced from (\ref{degenerate})--we 
have simple corollaries to the twofold-theorem of Szarek, Bengtsson and {\.Z}yczkowski 
\cite{sbz} of the form
\begin{equation}
\frac{P^{rank-4}_{C_{max} \in [0,1]}}{P^{rank-3}_{C_{max} \in [0,1]}} = 
2 -\frac{P^{rank-4}_{C_{max}=0}}{P^{rank-3}_{C_{max} \in [0,1]}} 
=2 -\frac{P^{rank-4}_{C_{max}=0}}{P^{rank-3}},
\end{equation}
where the $P$'s are Hilbert-Schmidt separability probabilities, for the real, complex 
or quaternionic two-qubit states.
\subsection{$C_{max} \in (0,\frac{1}{2}]$} \label{lowerpart}
So, the most conspicuous missing parts in the Hilbert-Schmidt
separability probability "puzzle" appear to us to be formulas for 
$P^{HS_{real}}_{C_{max} \in (0,\frac{1}{2}]}, 
P^{HS_{complex}}_{C_{max} \in (0,\frac{1}{2}]}$
and $P^{HS_{quat}}_{C_{max} \in (0,\frac{1}{2}]}$. Of course, 
we can subtract the sums of the other two parts 
($P^{HS_{real}}_{C_{max}=0}+ P^{HS_{real}}_{C_{max} \in [\frac{1}{2},1]}$, $P^{HS_{complex}}_{C_{max}=0} + P^{HS_{complex}}_{C_{max} \in [\frac{1}{2},1]}$ and $P^{HS_{quat}}_{C_{max}=0}+ P^{HS_{quat}}_{C_{max} \in [\frac{1}{2},1]}$)
 from our overall conjectures of $\frac{8}{17}, \frac{8}{33}$ 
and $\frac{72442944}{936239725}$ to obtain "induced" 
conjectures about these third components.

\begin{figure}
\includegraphics{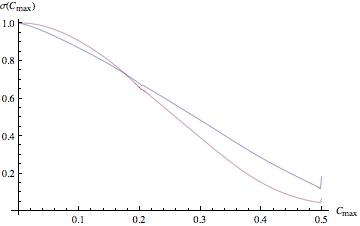}
\caption{\label{fig:Convenience}Previously-generated  
\cite[Fig. 1]{maxconcur4} estimates of 
the two-qubit real and complex eigenvalue-parameterized separability functions over the domain $C_{max} \in (0,\frac{1}{2}]$. The complex 
(red) curve is initially higher-valued and the real (blue) curve,  
close to linear.}
\end{figure}

\begin{figure}
\includegraphics{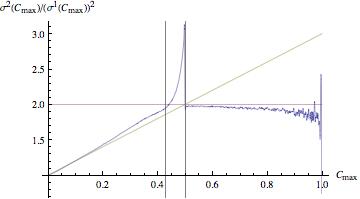}
\caption{\label{fig:Convenience2}Estimated ratio (for the full-rank two-qubit 
case) of 
$\frac{\sigma^2(C_{max}^{rank-4})}{(\sigma^1(C_{max}^{rank-4}))^2}$
along with the approximating lines  $1+2 C_{max}$ and 2 the 
vertical lines $C_{max}=\frac{43}{100}$ and $\frac{1}{2}$.}
\end{figure}

\begin{figure}
\includegraphics{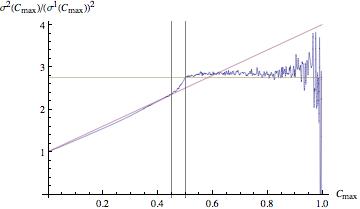}
\caption{\label{fig:ConvenienceArea}Estimated ratio (for the 
minimally-degenerate rank-three two-qubit 
case) of 
$\frac{\sigma^2(C_{max}^{rank-3})}{(\sigma^1(C_{max}^{rank-3}))^2}$
along with the approximating lines  $1+3 C_{max}$ and $\frac{11}{4}$ and the vertical lines 
$C_{max}=\frac{9}{20}$ and $\frac{1}{2}$.}
\end{figure}

Since it now appears crucial to, additionally, model
the eigenvalue-parameterized separability functions over the domain
$C_{max} \in (0,\frac{1}{2}]$, we present in Fig.~\ref{fig:Convenience}, 
for the convenience of the interested reader, the 
previously-generated \cite[Fig. 1]{maxconcur4} estimates of these functions. The real (blue) curve is close to linear ($\approx 1- 1.75
  C_{max}$). Also, we have noted that the complex (red) curve is 
quite well-fitted by $\cos^{22}{(C_{max})}$ and $\cos^{5}{(2 C_{max})}$. However, we appear here to lack a strictly similar Dyson-index ansatz to 
serve as a guide in constructing these two functions (cf. \cite{slater833}). Also, there were 
indications given in \cite[Figs. 3-5]{maxconcur4} that these two functions 
have multiple 
(matching) points of discontinuity in $C_{max} \in (0,\frac{1}{2}]$. 
(These were $C_{max} \approx 0.204, 0.294, 0.34$.)
The lack of a Dyson-index pattern strictly similar to that found apparently 
for $C_{max} \in [\frac{1}{2},1]$ to exploit for 
$C_{max} \in (0,\frac{1}{2}]$ is immediately apparent from 
Fig.~\ref{fig:Convenience2}. A 
{\it flat} line over at least some subdomain of $C_{max} \in (0,\frac{1}{2}]$ would indicate such a Dyson-index pattern. Clearly, no such flatness appears there. However, it
now seems that there is a pattern of the approximate form
\begin{equation} \label{pattern1}
\sigma^2(C_{max}^{rank-4}) = \Big( \sqrt{1+2 C_{max}^{rank-4}} 
\sigma^1(C_{max}^{rank-4}) \Big) ^2, \hspace{.1in} C_{max}^{rank-4} \in 
(0,\frac{1}{2}].
\end{equation}
(We have already noted that $\sigma^1(C_{max}^{rank-4}) 
\approx 1- 1.75 C_{max}^{rank-4}$ in this half-domain.) The highly interesting 
nature of Fig.~\ref{fig:Convenience2} led us to similarly
re-examine the minimally degenerate rank-3 two-qubit scenarios 
(Fig.~\ref{fig:maxconcurRank3}). Thus, we obtained 
Fig.~\ref{fig:ConvenienceArea}. Now $1+3 C_{max}$ serves as an 
excellent linear approximation, and we have 
a relation analogous to (\ref{pattern1})
\begin{equation} \label{pattern2}
\sigma^2(C_{max}^{rank-3}) = \Big( \sqrt{1+3 C_{max}^{rank-3}} 
\sigma^1(C_{max}^{rank-3}) \Big)^2, \hspace{.1in} C_{max}^{rank-3} \in 
(0,\frac{1}{2}].
\end{equation}
If we plot 
$\frac{\sigma^2(C_{max}^{rank-4})}{1+2 C_{max}^{rank-4}}$ 
{\it vs.} $(\sigma^1(C_{max}^{rank-4}))^2$ 
and also 
$\frac{\sigma^2(C_{max}^{rank-3})}{1+3 C_{max}^{rank-3}}$ 
{\it vs.} $(\sigma^1(C_{max}^{rank-3}))^2$ over the half-domain, the two curves within each
set are essentially indistinguishable.
We have investigated analogous plots  of the same form for the 
minimally-degenerate qubit-qutrit (Fig.~\ref{fig:maxconcurRank5}) and 
full-rank (Fig.~\ref{fig:maxconcurRank6}) 
cases over $C_{max} \in [0,\frac{1}{3}]$. They are much rougher in nature, due to our limited sampling, but still indicate initial monotonically-increasing (non-flat) behavior over $C_{max}$.
\subsection{{\it Total} probabilities over $C_{max}$ regions} \label{Total}
Let us also point out that the Hilbert-Schmidt probability 
($\tilde{P}$) that a generic  two-qubit real state (separable {\it or} entangled) lies in the 
domain $C_{max} \in [\frac{1}{2},1]$ is 
\begin{equation}
\tilde{P}^{HS_{real}}_{C_{max} \in [\frac{1}{2},1]}=
\frac{75962-25515 \sqrt{2} \tan ^{-1}\left(\sqrt{2}\right)}{2^{13} \cdot 3^3}
 \approx  0.187584,
\end{equation}
with the complex counterpart being
\begin{equation}
\tilde{P}^{HS_{complex}}_{C_{max} \in [\frac{1}{2},1]} = 
\frac{174957361466-124912178055 \sqrt{2} \cot
   ^{-1}\left(\frac{5}{\sqrt{2}}\right)}{2^{31} \cdot 3^5}
\approx 0.241961,
\end{equation}
and the quaternionic analogue being 
\begin{equation}
\tilde{P}^{HS_{quat}}_{C_{max} \in [\frac{1}{2},1]} 
=  \frac{\gamma_1- \gamma_2 \sqrt{2} \cot
   ^{-1}\left(\frac{5}{\sqrt{2}}\right)}{2^{63} \cdot 3^{10}} \approx  
0.323053,
\end{equation}
where
\begin{equation}
\gamma_1=217894901318574565900294,\hspace{.1in} \gamma_2=107614737772623370233945.
\end{equation}
(The comparable total probabilities for the minimally-degenerate
two-qubit states have been given in sec.~\ref{rankthreesub}.)
Since for the absolutely separable states ($C_{max}=0$), the two
probabilities 
$P_{C_{max} =0}$ and $\tilde{P}_{C_{max} =0}$ are equivalent, 
we can immediately determine (by subtracting  from 1 our known HS 
probabilities $P_{C_{max} =0}$ and 
$\tilde{P}_{C_{max} \in [\frac{1}{2},1]}$) the {\it complementary} probabilities,
$\tilde{P}_{C_{max} \in (0,\frac{1}{2}]}$.
Numerically, these are 
$\tilde{P}^{HS_{real}}_{C_{max} \in 
(0,\frac{1}{2}]} \approx 0.777582, \tilde{P}^{HS_{complex}}_{C_{max} \in 
(0,\frac{1}{2}]} \approx 0.754381$ and $\tilde{P}^{HS_{quat}}_{C_{max} \in 
(0,\frac{1}{2}]} \approx 0.676907$.

\section{Concluding remarks} \label{Remarks2}
Our analyses of two-qubit {\it diagonal-entry}-parameterized separability
functions (DESFs) \cite{slaterPRA2,slater833,slaterJGP2} and 
{\it eigenvalue}-parameterized separability functions (ESFs) 
\cite{JMP2008,maxconcur4} 
completely share a {\it common} goal: the determination of two-qubit separability
volumes and probabilities (in terms of various metrics). 
As pieces of these formidable objectives
begin to be assembled, we can pose a further challenge--to find
transformations between the two {\it different} 
sets of coordinates used--that is, 
(1)  the {\it diagonal entries} and (2) the {\it eigenvalues}
of $4 \times 4$ density matrices--that will map one set of 
separability functions
into the other.
The Schur-Horn
Theorem, which asserts
that the decreasingly-ordered vector of eigenvalues  of an
Hermitian matrix {\it majorizes} the decreasingly-ordered vector of
its diagonal entries \cite[chap. 4]{hornjohnson} (cf. \cite{nielsenvidal,carlen}), 
would appear to be of possible relevance in this regard, particularly 
since the maximal concurrence $C_{max}$ over spectral orbits (\ref{maxcon}) 
is expressed
in terms of the {\it ordered} eigenvalues.

In terms of the diagonal entries ($D_1,D_2,D_3,D_4=1-D_1-D_2-D_3$) 
of $4 \times 4$ two-qubit density matrices, we can express the conjectured
Hilbert-Schmidt separability probability \cite{slater833}, $\frac{8}{33}$,  of generic complex states in the form
\begin{equation} \label{desfconj}
\frac{8}{33} =\frac{12,108,096,000}{71} \int \int \int (D_1 D_2 D_3 D_4)^3 (3 -\nu)^2 \nu  d D_1 d D_2 d D_3,
\end{equation}
where $\nu \equiv 
\frac{D_1 D_4}{D_2 D_3}$, and the integration extends over
the unit simplex, but with the restriction $\nu \leq 1$. (We note that 
$12,108,096,000=2^9 \cdot 3^3 \cdot 5^3 \cdot 7^2 \cdot 11 \cdot 13$. Let us also observe that the variable $\log{\nu}$ conveniently ranges over the entire real axis and is symmetric about the origin.)

Additionally, in terms of the eigenvalues ($\lambda_1,\lambda_2,\lambda_3,1-\lambda_1-\lambda_2-\lambda_3$) of $4 \times 4$ two-qubit density matrices, we can express this {\it same}
separability probability as (cf. (\ref{ansatz}))
\begin{equation} \label{esfconj}
\frac{8}{33} = 2,201,472,000 
\int \int \int \sigma^{(2)}(C_{max}(\lambda_1\ldots\lambda_4)) 
\Pi_{i<j}^4 (\lambda_i-\lambda_j)^2  d \lambda_1 d \lambda_2  d \lambda_3,
\end{equation}
and the integration extends over that part (Weyl chamber \cite{ingemarkarol}) of the unit simplex for which 
$\lambda_1 \geq \lambda_2 \geq \lambda_3 \geq \lambda_4$. 
(We note that, interestingly, in light of the just previous factorization, 
$2,201,472,000=2^{10} \cdot 3^3 \cdot 5^3 \cdot 7^2 \cdot 13$.) Here $\sigma^{(2)}(C_{max}(\lambda_1\ldots\lambda_4))$ is the (two-qubit complex 
[$\beta=2$]) eigenvalue-parameterized separability function that we have previously sought to determine \cite[Fig. 1]{maxconcur4}, and was found to be very well-fitted  by
$\frac{(2-2 C_{max})^3}{15}$ for $C_{max} \in [\frac{1}{2},1]$ 
(Fig.~\ref{fig:normed2} and (\ref{ratio2})). (It is possible to reexpress these two last integrals so that {\it both} are taken over the 
{\it same} complete 3-dimensional unit simplex.) 
Further still, our generic complex 
two-qubit {\it Bures} separability probability conjecture (\ref{silvermean}) 
\cite[Table VI]{slaterJGP} takes the form
\begin{equation} \label{esfconjBures}
\frac{1680 (\sqrt{2}-1)}{\pi^8} = \frac{\pi^2}{71680}
\int \int \int \frac{\sigma^{(2)}(C_{max}(\lambda_1\ldots\lambda_4))}{\sqrt{\lambda_1 \lambda_2 \lambda_3 \lambda_4}} 
\Pi_{i<j}^4 \frac{(\lambda_i-\lambda_j)^2}{\lambda_i + \lambda_j} 
d \lambda_1 d \lambda_2  d \lambda_3.
\end{equation}

It is abundantly clear: (a) 
that this (piecewise continuous) function 
$\sigma^{(2)}(C_{max}(\lambda_1\ldots\lambda_4))$ 
has a jump discontinuity at $C_{max}=\frac{1}{2}$ (as well as does its 
real counterpart $\sigma^{(1)}(C_{max}(\lambda_1\ldots\lambda_4))$; 
and (b) that
in the diagonal-entry-parameterized scenario, the value 
$\nu=\frac{D_1 D_4}{D_2 D_3}=1$ 
is a locus of special symmetry. In this regard, we might speculate that if one can find a coordinate transformation between the two separability probability expressions ((\ref{desfconj}) and (\ref{esfconj})), then those values of the $\lambda_i$'s for which $C_{max}=\frac{1}{2}$ will be mapped to those values of the $D_i$'s for which $\nu=1$.

\begin{figure}
\includegraphics{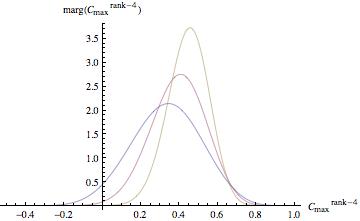}
\caption{\label{fig:Notexplicit}Full-rank two-qubit univariate marginal probability distributions over $C_{max}^{rank-4} \in 
[-\frac{1}{2},1]$ of the three-dimensional Hilbert-Schmidt measures on the eigenvalues. The quaternionic 
curve has the highest peak, and the real curve, the lowest. Numerical 
methods were used.}
\end{figure}

\begin{figure}
\includegraphics{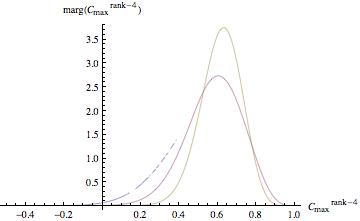}
\caption{\label{fig:NotexplicitBures}Counterpart to 
Fig.~\ref{fig:Notexplicit} based on the Bures metric. The quaternionic 
curve has the highest peak, followed by the complex curve, with the (incompletely constructed) real curve,  apparently the lowest. Numerical methods were used.}
\end{figure}

Through the use of the jacobian transformation of the diagonal entry $D_3$ (say) to $\nu$ 
\cite[eq. (11)]{slaterPRA2}, and subsequent integration over $D_1$ and 
$D_2$, it is possible to explicitly reduce the computation of the trivariate
integral (\ref{desfconj}) to that of 
a univariate integral in $\nu$. 
In Fig.~\ref{fig:Notexplicit}, we show--based on numerical calculations--the univariate marginal probability distributions of the Hilbert-Schmidt measure over the real and complex two-qubit states in terms of $C_{max}^{rank-4}$ (cf. (\ref{conctransf1})-(\ref{conctransf3})). 
Similarly to their rank-3 counterparts( Fig.~\ref{fig:marginalConc}), these curves have differently-positioned peaks, and are not symmetric, but skewed to the right.
(We take the range of  $C_{max}^{rank-4}$ to be $[-\frac{1}{2},1]$, 
to accord with actual values, rather than the conventional 
$[0,1]$ [cf. (\ref{maxcon})].) In Fig.~\ref{fig:NotexplicitBures}, we show
the  Bures-metric counterpart, although
we encounter some "glitch" in displaying the real curve here.

The counterparts to the formulas (\ref{desfconj}) and (\ref{esfconj}), in light of our
conjecture \cite{slater833} that the Hilbert-Schmidt separability probability of the generic {\it real} two-qubit states is $\frac{8}{17}$,  
are (the domains of integration being the same)
\begin{equation} \label{desfconjreal}
\frac{8}{17} =\frac{1,209,600}{17} \int \int \int (D_1 D_2 D_3 D_4)^{3/2} (3 -\nu) \sqrt{\nu} d D_1 d D_2 d D_3,
\end{equation}
and
\begin{equation} \label{esfconjreal}
\frac{8}{17} =\frac{15,482,880}{17}
\int \int \int \sigma^{(1)}(C_{max}(\lambda_1\ldots\lambda_4)) 
\Pi_{i<j}^4 (\lambda_i-\lambda_j)  d \lambda_1 d \lambda_2  d \lambda_3.
\end{equation}
(Here, $1,209,600=2^8 \cdot 3^3 \cdot 5^2 \cdot 7$ and $15,482,880=2^{14} \cdot 3^3 \cdot 5 \cdot 7$. It appears 
[Fig.~\ref{fig:Rank3counterpart} and (\ref{ratio2})) that possibly 
$\sigma^{(1)}(C_{max}(\lambda_1\ldots\lambda_4))
 = \frac{(2-2 C_{max})^{\frac{3}{2}}}{\sqrt{30}}$ for $C_{max} \in [\frac{1}{2},1]$.)

Let us point out the possible relevance of the concept
of the {\it Thouless energy} \cite[p. 734]{beenakker} in the modeling of the threshold or crossover effect
we have numerically observed for eigenvalue-parameterized separability functions in both the full generic real and complex 
two-qubit and qubit-qutrit cases. There, the Dyson indices ($\beta=1, 2$) 
of random matrix theory only seemed to apply above a certain value of 
the maximal concurrence $C_{max}$ (that is, $\frac{1}{2}$ in the two-qubit case, and possibly $\frac{1}{3}$ in the qubit-qutrit instance). (It remains to formally reconcile these observations with the ones that, in terms of diagonal-entry-parameterized separability functions, Dyson-index behavior appear to be {\it strictly} followed \cite{slater833}.)

\begin{acknowledgments}
I would like to express appreciation to the Kavli Institute for Theoretical
Physics (KITP)
for computational support in this research, as well as to 
K. {\.Z}yczkowski for his interest and 
for a number of suggestions concerning the analyses and their presentation.
\end{acknowledgments}

\bibliography{GPHrevise2}

\end{document}